\documentclass[lettersize,journal]{IEEEtran}
\usepackage{cite}
\usepackage{amsmath,amssymb,amsfonts}
\usepackage{algorithmic}
\usepackage{algorithm}
\usepackage{array}
\usepackage[caption=false,font=normalsize,labelfont=sf,textfont=sf]{subfig}
\usepackage{textcomp}
\usepackage{stfloats}   
\usepackage{graphicx}
\usepackage{bm}
\usepackage{cases}
\usepackage{xcolor}
\usepackage{url}
\newtheorem{theorem}{Theorem}[section]
\newtheorem{lemma}{Lemma}[section]
\usepackage{verbatim}
\usepackage{graphicx}
\usepackage{cite}
\usepackage{booktabs}
\usepackage{hyperref}
\begin{document}

\title{Enhancing Physical Layer Security in MIMO Systems Assisted by Beyond-Diagonal Reconfigurable Intelligent Surfaces}

\author{Weijie Xiong, Jingran Lin, Cunhua Pan, Yilong Zeng, and Qiang Li
\thanks{This work was supported by the Natural Science Foundation of China under Grant 62171110. \textit{(Corresponding author: Jingran Lin)}.}
\thanks{Jingran Lin and Qiang Li are with the School of Information and Communication Engineering, University of Electronic Science and Technology of China, Chengdu 611731, China, the Laboratory of Electromagnetic Space Cognition and Intelligent Control, Beijing 100083, China, and also with the Tianfu Jiangxi Laboratory, Chengdu, Sichuan 641419, China (e-mail: jingranlin@uestc.edu.cn; lq@uestc.edu.cn).}
\thanks{Weijie Xiong and Yilong Zeng are with the School of Information and Communication Engineering, University of Electronic Science and Technology of China, Chengdu 611731, China (e-mail: 202311012313@std.uestc.edu.cn; 202321010815@std.uestc.edu.cn)}\thanks{Cunhua Pan is with the National Mobile Communications Research Laboratory, Southeast University, China (e-mail: cpan@seu.edu.cn)}.
}

\markboth{Journal of \LaTeX\ Class Files,~Vol.~14, No.~8, August~2021}%
{Shell \MakeLowercase{\textit{et al.}}: A Sample Article Using IEEEtran.cls for IEEE Journals}


\maketitle

\begin{abstract}
Reconfigurable intelligent surfaces (RISs) hold significant promise for enhancing physical layer security (PLS). However, conventional RISs are typically modeled using diagonal scattering matrices, capturing only independent reflections from each reflecting element, which limits their flexibility in channel manipulation. In contrast, beyond-diagonal RISs (BD-RISs) employ non-diagonal scattering matrices enabled by active and tunable inter-element connections through a shared impedance network. This architecture significantly enhances channel shaping capabilities, creating new opportunities for advanced PLS techniques. This paper investigates PLS in a multiple-input multiple-output (MIMO) system assisted by BD-RISs, where a multi-antenna transmitter sends confidential information to a multi-antenna legitimate user while a multi-antenna eavesdropper attempts interception. To maximize the secrecy rate (SR), we formulate it as a non-convex optimization problem by jointly optimizing the transmit beamforming and BD-RIS REs under power and structural constraints. To solve this problem, we first introduce an auxiliary variable to decouple BD-RIS constraints. We then propose a low-complexity penalty product Riemannian conjugate gradient descent (P-PRCGD) method, which combines the augmented Lagrangian (AL) approach with the product manifold gradient descent (PMGD) method to obtain a Karush-Kuhn-Tucker (KKT) solution. Simulation results confirm that BD-RIS-assisted systems significantly outperform conventional RIS-assisted systems in PLS performance.
\end{abstract}


\section{Introduction}
\IEEEPARstart{W}{ith} the significant increase in network heterogeneity, concerns regarding the security of next-generation wireless communication systems have become more pronounced \cite{chorti2022context}. Over the past few decades, cryptographic encryption and decryption have been the primary methods for ensuring information security. However, due to the high computational complexity and key management overhead associated with these techniques, there has been growing interest in achieving or enhancing information security at the physical layer, a concept known as physical layer security (PLS) \cite{mitev2023physical}.

One key performance metric of PLS is the secrecy rate (SR) \cite{bereyhi2021securing}, and maximizing SR in the multiple-input multiple-output (MIMO) communication systems has attracted significant attention over the past decade \cite{tubail2023physical}. Generally, SR is characterized by the channel capacity difference between the legitimate user and the eavesdropper. Popular PLS techniques include transmit beamforming \cite{zhu2014secure,zhu2015linear,wu2016secure} and artificial noise (AN)-aided designs \cite{yang2023secure,nguyen2016joint}, which exploit channel characteristic differences to enhance the reception quality of legitimate users or degrade that of eavesdroppers. For instance, \cite{zhu2014secure} first analyzed PLS in a multicell MIMO system using maximum ratio transmission (MRT) beamforming and null-space AN under both perfect and imperfect channel state information (CSI). Building on this, \cite{zhu2015linear} proposed an advanced beamforming design based on matrix polynomials. The study in \cite{wu2016secure} examined the impact of active eavesdropping in MIMO systems and derived SR using matched filter beamforming. To reduce the hardware cost, \cite{yang2023secure} further investigated the use of hybrid beamforming in MIMO systems with AN. However, these approaches assume that both the legitimate user and the eavesdropper experience spatially diverse channels, which may not always be satisfied in practical scenarios \cite{tang2020wireless}. For instance, when the legitimate user and the eavesdropper are in close proximity, their channels can become highly correlated, leading to significant performance degradation \cite{lin2017physical}.

To address this challenge, reconfigurable intelligent surfaces (RISs) have gained significant attention in recent years \cite{tang2020wireless,xiong2025constant}. An RIS is a software-controlled metasurface equipped with passive, digitally controlled reflection elements (REs), enabling it to intelligently reconfigure the wireless propagation environment \cite{wu2021intelligent}. When applied to PLS, RIS can proactively suppress the correlation between legitimate and eavesdropping channels by dynamically optimizing its REs, while simultaneously enhancing the reception quality of the legitimate user through additional reflected signals \cite{khoshafa2024ris}. Inspired by this capability, numerous studies have explored enhancing PLS performance in MIMO systems with RIS assistance \cite{asaad2022secure,hong2020artificial,dong2020enhancing,yu2020robust,wang2022intelligent,cui2019secure}. For instance, \cite{asaad2022secure} investigates the SR achieved through joint beamforming and REs optimization at the transmitter and RIS in a MIMO system. To further enhance performance, \cite{hong2020artificial} and \cite{dong2020enhancing} proposes AN-aided joint transmission schemes for RIS-assisted MIMO systems, specifically targeting improved PLS in the presence of passive attacks. In more practical scenarios, the SR of systems with non-colluding and colluding eavesdroppers is maximized by jointly optimizing the beamforming and REs, as proposed in \cite{yu2020robust} and \cite{wang2022intelligent}, respectively. Additionally, \cite{cui2019secure} examines systems with perfect CSI to mitigate passive attacks when the eavesdropping channels are stronger than the legitimate channels.

Notably, previous studies primarily focus on the diagonal RIS (D-RIS), where each RE operates independently and is connected solely to its self-impedance \cite{shen2021modeling}. This configuration restricts the REs to a diagonal matrix, thereby reducing their effectiveness in more complex environments \cite{shen2021modeling}. For instance, with the continuous advancement of eavesdropping techniques such as collaborative eavesdropping \cite{huang2021navigation}, UAV-assisted interception \cite{dai2022unmanned}, and near-field eavesdropping \cite{zhang2024physical}, the spatial correlation of wireless transmissions is increasingly compromised, rendering traditional D-RIS approaches insufficient against these emerging threats. Therefore, it is necessary to further explore and improve the structure and working principles of D-RIS to address these challenges. To overcome these limitations and achieve greater flexibility, a promising evolution known as beyond-diagonal RIS (BD-RIS) has recently been introduced \cite{li2023dynamic}. In this architecture, each RIS RE is connected not only to its self-impedance but also to other elements through internal impedances, leading to two configurations: fully-connected BD-RIS and group-connected BD-RIS \cite{li2023reconfigurable}. In a fully-connected BD-RIS, all REs are interconnected, resulting in a full matrix. Conversely, in a group-connected BD-RIS, subgroups of REs are interconnected, leading to a block-diagonal matrix. Similar to D-RIS, BD-RIS requires only low-power, low-cost components to enable the reconfigurability of its REs \cite{shen2021modeling}.

With its enhanced flexibility, BD-RIS offers finer control over the wireless environment, significantly improving coverage and spectral efficiency. In particular, several studies have investigated the use of BD-RIS to enhance performance in MIMO systems \cite{shen2021modeling,li2023dynamic,fang2023low,nerini2023closed,nerini2023discrete,santamaria2023snr,nerini2023pareto}. For example, the pioneering work in \cite{shen2021modeling} examined the scaling law of received signal power for both fully-connected and group-connected BD-RISs, demonstrating that these configurations significantly improve power reflection efficiency, outperforming conventional D-RIS. In \cite{li2023dynamic}, the authors introduced a novel group-connected BD-RIS with a dynamic grouping strategy, wherein REs were adaptively divided into subsets based on the perfect CSI of users within a MIMO system. Low-complexity closed-form optimization strategies for BD-RISs focusing on power minimization in MIMO systems were proposed in \cite{fang2023low} and \cite{nerini2023closed}, while the impact of discrete coefficients on BD-RIS performance was studied in \cite{nerini2023discrete}. Additionally, \cite{santamaria2023snr} explored the signal-to-noise ratio (SNR) maximization problem for MIMO channels assisted by BD-RISs, and \cite{nerini2023pareto} derived the Pareto frontier for the performance-complexity trade-off of different BD-RIS configurations. 

Building on its success in general wireless communications, the potential of BD-RIS to enhance PLS has also been explored \cite{agarwal2023enhanced,lin2024securing,khan2025enhancing}. The work in \cite{agarwal2023enhanced} pioneered this direction by demonstrating the effectiveness of BD-RIS in securing wireless transmissions. More recently, this concept has been extended to emerging systems, including those enabled by UAVs \cite{lin2024securing} and cognitive radio networks \cite{khan2025enhancing}. Despite these promising developments, existing studies are limited to multiple-input single-output (MISO) wiretap channels. As a result, the performance gains and practical applicability of BD-RIS in more general and challenging MIMO wiretap scenarios remain unestablished, representing the primary research gap this paper aims to address.

Motivated by the demonstrated benefits of BD-RIS in the MISO case and the need to counter advanced eavesdropping threats in MIMO systems, this paper investigates the PLS of a BD-RIS-assisted MIMO system. Specifically, we consider a scenario where a multi-antenna transmitter sends confidential information to a multi-antenna legitimate user with the assistance of the BD-RIS, while a multi-antenna eavesdropper attempt to intercept the transmission. The main contributions of this work are summarized as follows:
\begin{itemize}
\item This paper investigates PLS in BD-RIS-assisted MIMO systems, extending prior work that was limited to MISO scenarios. Unlike the well-established case of D-RIS in MIMO systems \cite{asaad2022secure,hong2020artificial,dong2020enhancing,yu2020robust,wang2022intelligent,cui2019secure}, the BD-RIS architecture imposes challenging non-convex orthogonality constraints and variable coupling due to its unique impedance network. These complexities render existing D-RIS-based algorithms unsuitable and substantially increase the optimization difficulty.
\item We formulate the problem as a non-convex optimization task aimed at maximizing the SR by jointly optimizing the transmit beamforming at the transmitter and REs at the BD-RIS, while ensuring compliance with total transmit power constraints and the structural properties of BD-RIS.
\item To address the problem, we reformulate it by introducing an auxiliary variable to decouple the BD-RIS constraints. We then propose a penalty product Riemannian conjugate gradient descent (P-PRCGD) method that combines the augmented Lagrangian (AL) method and the product manifold gradient descent (PMGD) algorithm, ensuring convergence to a Karush-Kuhn-Tucker (KKT) point. Specifically, the AL method relaxes the equality constraints of the auxiliary variable into the objective function, transforming the problem into an unconstrained optimization over a product Riemannian manifold (PRM). The P-PRCGD algorithm is employed to efficiently solve the problem in this manifold space.
\item The SR maximization problem is also extended to a more practical imperfect CSI scenario that incorporates channel estimation errors (CEEs). We demonstrate that the proposed P-PRCGD method can be adapted to solve this problem efficiently with only a minor modification.
\end{itemize}

The remainder of this paper is organized as follows. Section II introduces the system model and formulates the optimization problem. Section III presents the proposed P-PRCGD framework for solving the problem. In Section IV, this framework is extended to the imperfect CSI scenario. Section V provides numerical results to validate our approach, and Section VI concludes the paper.

The following notations are used in the paper. Vectors and matrices are denoted by $\mathbf{a}$ and $\mathbf{A}$, respectively. The operators $(\cdot)^T$, $(\cdot)^H$, and $(\cdot)^*$ represent the transpose, conjugate transpose, and conjugate of a matrix or vector. $\mathbf{I}$ denotes the identity matrix, and ${\mathbb{C}}^N$ represents the space of complex vectors of dimension $N$. The complex Gaussian distribution with mean $\mu$ and variance $\sigma^2$ is $\mathcal{CN}(\mu, \sigma^2)$. The operators $\text{Tr}(\cdot)$, $\|\cdot\|$, and $|\cdot|$ denote the trace, Frobenius norm, and absolute value, respectively. The phase and real part of $\mathbf{A}$ are given by $\text{arg}(\mathbf{A})$ and $\Re(\mathbf{A})$, respectively. The Hadamard division is given by $\mathbf{A} \oslash \mathbf{B}$, and $\text{blkdiag}(\mathbf{A}, \mathbf{B})$ represents a block diagonal matrix formed by placing $\mathbf{A}$ in the top-left block and $\mathbf{B}$ in the bottom-right block, with all off-diagonal blocks being zero. The direct sum operation satisfies $\mathbf{A} \oplus \mathbf{B} = \text{blkdiag}(\mathbf{A}, \mathbf{B})$, and $\langle \mathbf{A}, \mathbf{B} \rangle$ denotes the inner product.

\section{System Model and Problem Formulation}

\subsection{Signal Model}
\begin{figure}[htbp]
  \begin{center}
  \includegraphics[width=2.5in]{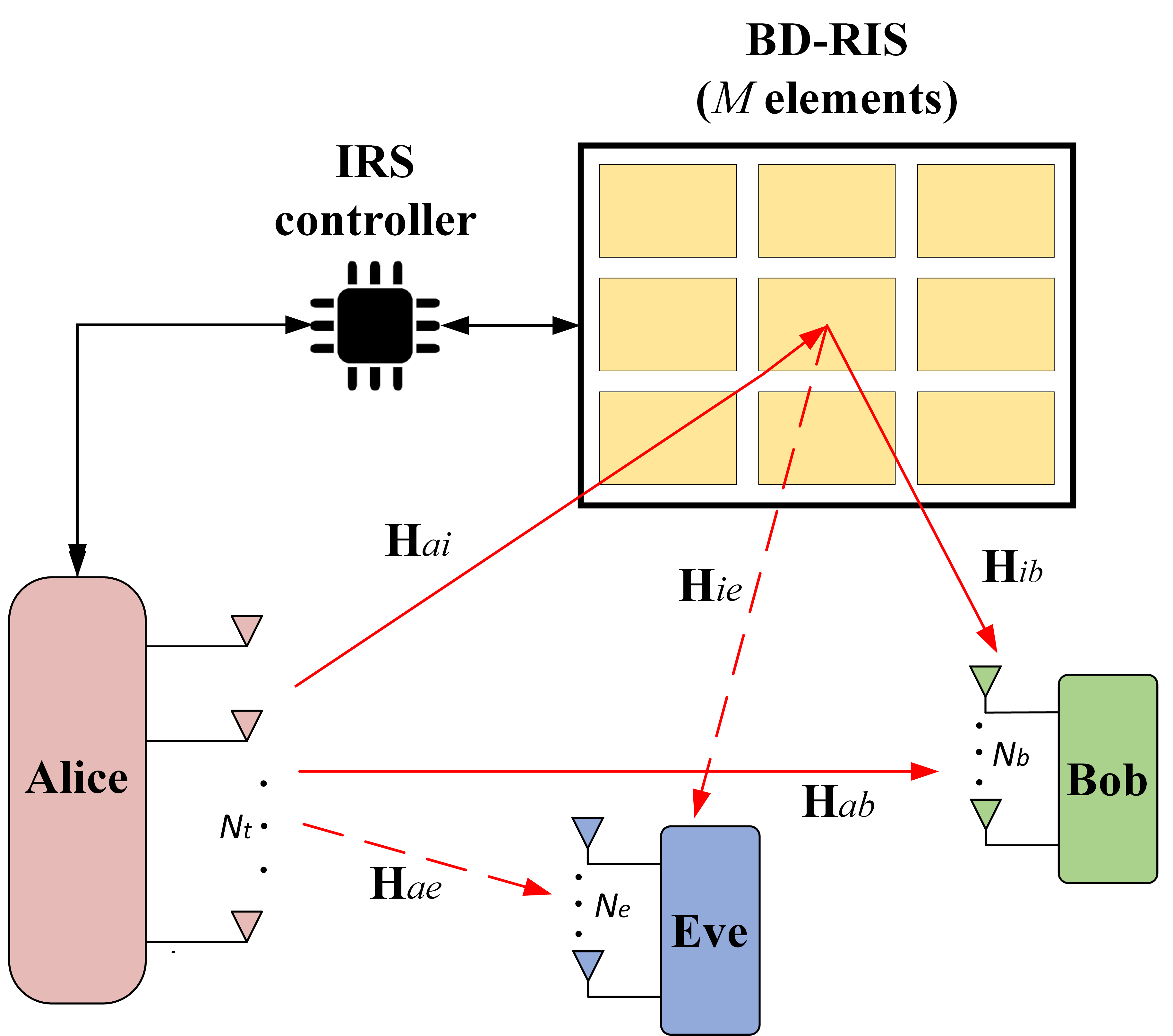}\\
  \caption{Illustration of the MIMO downlink network with a BD-RIS.}\label{1}
  \end{center}
\end{figure}

Consider a MIMO downlink network assisted by a BD-RIS, as shown in Fig. \ref{1}, where the BD-RIS is deployed to assist legitimate communication between the transmitter (Alice) and the legitimate user (Bob) in the presence of an eavesdropper (Eve), which attempts to intercept information from the communication. The number of antennas at Alice, Bob, and Eve are \(N_t\), \(N_b\), and \(N_e\), respectively, while the BD-RIS consists of \(M\) REs. 

Let \( {\bf s} \in \mathbb{C}^{N_s} \) represent the confidential information for Bob, the transmitted signal $\bf x$ from Alice can be expressed as,
\begin{equation}
{\bf x} =  {\bf W} {\bf s} \in \mathbb{C}^{N_t},\label{transmitsignal}
\end{equation}
where \( \mathbb{E}[{\bf s}{\bf s}^H] = {\bf I}_{N_s} \); \( N_s \leq \min(N_t, N_b) \) denotes the number of data streams observed by Bob; \( {\bf W} \in \mathbb{C}^{N_t \times N_s} \) is the transmit beamformer. In this paper, we assume full transmit power is used for secure communication, i.e., $\|{\bf W}\|^2 = P$, which is reasonable for SR maximization; see \cite{5485016, zhou2025joint} for details.

Denote \( \mathbf{H}_{ab} \in \mathbb{C}^{N_b \times N_t} \), \( \mathbf{H}_{ae} \in \mathbb{C}^{N_e \times N_t} \), \( \mathbf{H}_{ai} \in \mathbb{C}^{M \times N_t} \), \( \mathbf{H}_{ib} \in \mathbb{C}^{N_b \times M} \), and \( \mathbf{H}_{ie} \in \mathbb{C}^{N_e \times M} \) as the channels between Alice-Bob, Alice-Eve, Alice-RIS, RIS-Bob, and RIS-Eve, respectively. Assuming quasi-static channels, the received signals at Bob and Eve are expressed as,
\begin{equation}
\begin{aligned}
& \mathbf{y}_b = ( \mathbf{H}_{ib} {\bf{\Theta}} \mathbf{H}_{ai} + \mathbf{H}_{ab} ) \mathbf{x}+ {\bf{n}}_b \in \mathbb{C}^{N_b}, \\
& \mathbf{y}_e = ( \mathbf{H}_{ie} {\bf{\Theta}} \mathbf{H}_{ai} + \mathbf{H}_{ae}  )\mathbf{x} + {\bf{n}}_e \in \mathbb{C}^{N_e},\label{recievesignal}
\end{aligned}
\end{equation}
where \( {\bf{\Theta}} \in \mathbb{C}^{M \times M} \) denotes the RE matrix for the BD-RIS; \( {\bf{n}}_b \sim \mathcal{CN}(\mathbf{0}, \sigma_b^2\mathbf{I}_b) \) and \( {\bf{n}}_e \sim \mathcal{CN}(\mathbf{0}, \sigma_e^2\mathbf{I}_e) \) represent the vectors of zero-mean
additive white Gaussian noise (AWGN).

Based on (\ref{transmitsignal}) and (\ref{recievesignal}), SR from Alice to Bob is given by, 
\begin{equation}
R_{sec}({\bf W},{\bm \Theta}) = [R_b({\bf W},{\bm \Theta}) - R_e({\bf W},{\bm \Theta})]^+,\label{Defiser}
\end{equation}  
where \([z]^+ = \max(0, z)\); \( R_b({\bf W},{\bm \Theta}) \) and \( R_e({\bf W},{\bm \Theta}) \) denote the channel capacities of Bob and Eve, respectively, given by,
\begin{subequations}
\begin{align}
& {R_b({\bf W},{\bm \Theta})}= \log \left|\sigma_b^2{\bf I}_{N_b}+{\bf H}_b({\bm \Theta}){\bf W}{\bf W}^H{\bf H}_b^H({\bm \Theta})  \right|, \\
& {R_e({\bf W},{\bm \Theta})}= \log \left|\sigma_e^2{\bf I}_{N_e}+{\bf H}_e({\bm \Theta}){\bf W}{\bf W}^H{\bf H}_e^H({\bm \Theta})  \right|,
\end{align}
\label{trasnmittr}%
\end{subequations}
where,
\begin{equation}
\begin{aligned}
    &{\bf H}_b({\bm \Theta})=\mathbf{H}_{ib} {\bf{\Theta }} \mathbf{H}_{ai} + \mathbf{H}_{ab}\in \mathbb{C}^{N_b \times N_t},\\ 
    &{\bf H}_e({\bm \Theta})=\mathbf{H}_{ie} {\bf{\Theta }} \mathbf{H}_{ai} + \mathbf{H}_{ae}\in \mathbb{C}^{N_e \times N_t}.
\end{aligned}
\end{equation}

\textit{\textbf{Remark 1:}} In this paper, we assume that CSI of Eve is available to Alice. This can be achieved through various methods, such as the CSI feedback method or even the local oscillator power leakage from Eve's RF frontend \cite{mukherjee2012detecting}. Additionally, we initially assume that the transmitter has perfect CSI of Eve to establish a performance upper bound. In Section IV, we extend our design to account for imperfect CSI.

\textit{\textbf{Remark 2:}} Our analysis adopts the widely used idealized channel model for BD-RIS-assisted networks \cite{nerini2024universal}, which assumes perfect impedance matching, no mutual coupling, and relies on the unilateral approximation. While a fully physics-compliant model would account for additional hardware-level non-idealities, this established signal processing framework remains essential for deriving fundamental insights and theoretical performance bounds \cite{nerini2023discrete, santamaria2023snr, nerini2023pareto}. For a detailed comparison with a fully physics-compliant model, readers are referred to \cite{del2025physics}.

\subsection{RIS Architectures}
\begin{figure}[htbp]
  \begin{center}
  \includegraphics[width=3.5in]{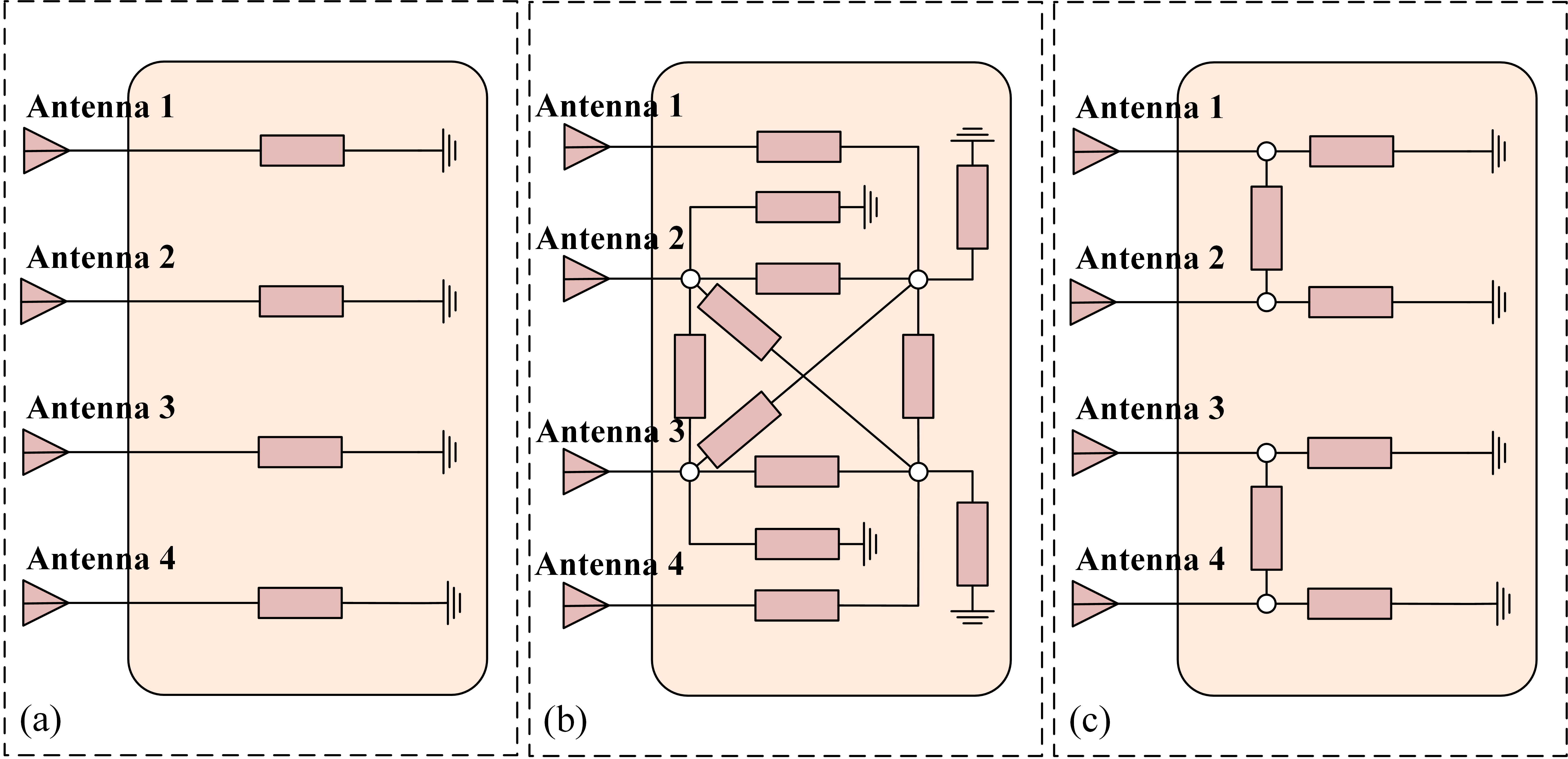}\\
  \caption{Illustration of circuit network topologies for a 4-antenna RIS with (a) D-RIS, (b) full-connected BD-RIS, and (c) group-connected BD-RIS.}\label{2}
  \end{center}
\end{figure}

Depending on the circuit network topology, RIS architectures can be classified into three categories \cite{shen2021modeling}: (a) D-RIS, (b) fully-connected BD-RIS, and (c) group-connected BD-RIS, as illustrated in Fig. \ref{2} for the specific case where \( M = 4 \). In this paper, we focus on BD-RIS architectures, specifically (b) and (c). Details are provided below.

\subsubsection{Fully-connected BD-RIS}
This circuit network topology involves all REs being connected to each other through reconfigurable impedance components. In Fig. \ref{2}(b), we provide an example of a fully-connected BD-RIS. Thus, ${\bm \Theta}$ is a full matrix that satisfies the following constraints,
\begin{equation}
    {\bm \Theta} = {\bm \Theta}^T, {\bm \Theta}{\bm \Theta}^H={\bf I}_M, \label{fcr}
\end{equation}
Clearly, this architecture can achieve the best performance due to the greatest DoFs it provides.

\subsubsection{Group-connected BD-RIS}
It is evident that as the number \( M \) increases, the circuit topology of the fully-connected BD-RIS becomes increasingly complex. To achieve a balance between RIS performance and circuit complexity, the group-connected topology is proposed. In the group-connected BD-RIS, the \( M \) elements are divided into \( G \) groups, indexed by \( \mathcal{G} = \{1, \dots, G\} \), with each group containing an equal number of elements \( \tilde{M} = \frac{M}{G} \). An example of a BD-RIS with a group-connected architecture having 2 groups is illustrated in Fig. \ref{2}(c). Thus, \( {\bf \Theta} \) is a block-diagonal matrix, which is given by,
\begin{equation}
    \begin{aligned}
       & {\bf \Theta} = \text{blkdiag}({\bf \Theta}_1,...,{\bf \Theta}_G),\\
       & {\bm \Theta}_g = {\bm \Theta}_g^T\in\mathbb{C}^{{\tilde M}\times{\tilde M}}, {\bm \Theta}_g{\bm \Theta}_g^H={\bf I}_{\tilde M}, \forall g \in \mathcal{G}.\label{gcbdris}
    \end{aligned}
\end{equation}

In summary, compared to the conventional D-RIS depicted in Fig. \ref{2}(a), the BD-RIS architectures shown in Fig. \ref{2}(b) and Fig. \ref{2}(c) connect each RE not only to its own impedance but also to other elements through internal impedances. These configurations provide beyond-diagonal matrices for REs with additional DoFs, offering opportunities to enhance PLS.

\textit{\textbf{Remark 3:}} We should mention that the equations (\ref{fcr}) and (\ref{gcbdris}) are valid only under the assumptions of a reciprocal and lossless BD-RIS. For detailed derivations, readers are encouraged to refer to \textit{Appendix A and B} in \cite{li2025tutorial}. In practice, non-idealities such as quantization errors, phase noise, and hardware impairments in RIS elements affect the secrecy rate improvements \cite{alexandropoulos2021reconfigurable}. However, for simplicity and to isolate the core performance gains of the BD-RIS architecture, these factors are not considered in this analysis. Since the problem is already challenging under the ideal model, extending the algorithm to account for practical constraints is left for future work.

\subsection{Problem Formulation}
In this paper, we aim to maximize SR by jointly optimizing the transmit beamforming matrix \( \mathbf{W} \) at Alice and the RE matrix \( \mathbf{\Theta} \) at the BD-RIS, subject to constraints on transmit power and the structure of the BD-RIS. In particular, we focus on the formulation by considering the group-connected BD-RIS, as it can also be tended to the fully-connected RIS by setting \( G = 1 \) . Therefore, we formulate the SR maximization problem as follows, 
\begin{subequations}
\begin{align}
& \max _{{\bf W}, {\bm \Theta}} \quad R_{sec}({\bf W},{\bm \Theta}), \label{objorig}\\
& \text { s.t. }\quad \left\|{\bf W}\right\|^2=P, \label{cons1orig}\\
 &\quad\quad\quad {\bf \Theta} = \text{blkdiag}({\bf \Theta}_1,...,{\bf \Theta}_G),\label{cons2orig}\\
       &\quad\quad\quad {\bm \Theta}_g = {\bm \Theta}_g^T, {\bm \Theta}_g{\bm \Theta}_g^H={\bf I}_{\tilde M}, \forall g \in \mathcal{G}. \label{endorig}
\end{align}
\label{overallproblem}%
\end{subequations}
The primary difficulty in solving problem (\ref{overallproblem}) lies with the BD-RIS, which introduces complex non-convex unitary constraints (\ref{endorig}) and couples the optimization variables between the objective function (\ref{objorig}) and the constraint (\ref{endorig}). Without the BD-RIS, this problem reduces to the classical secrecy capacity maximization of a MIMO wiretap channel, for which efficient algorithms are well-established \cite{loyka2015algorithm}. However, the complexities introduced by the BD-RIS render existing methods inapplicable. Accordingly, this paper develops a tractable approach to find a high-quality solution to problem (\ref{overallproblem}).

\textit{\textbf{Remark 4:}} It is important to emphasize that the SR maximization problem in (\ref{overallproblem}) is formulated based on the idealized assumptions in \textit{\textbf{Remarks 1–3}}. Although this formulation differs from a physics-compliant model \cite{tapie2025beyond}, the idealized model, along with its corresponding efficient algorithm, facilitates the exploration of signal processing characteristics and the performance limits that BD-RIS can offer for secure communications.

\section{Secrecy Rate Maximization Framework for BD-RIS-Assisted MIMO Systems}
In this section, we propose a framework for the SRM problem (\ref{overallproblem}). Firstly, we first reformulate the problem (\ref{overallproblem}) into a more manageable form. Then, we introduce the P-PRCGD method to efficiently solve the reformulated problem.

\subsection{Problem Reformulation}
To begin with, the problem (\ref{overallproblem}) is equivalently written as (after dropping the log operation),
\begin{equation}
\begin{aligned}
\max _{{\bf W}, {\bm \Theta}} \quad & \max \left\{  \frac{\left|\sigma_b^2{\bf I}_{N_b}+{\bf H}_b({\bm \Theta}){\bf W}{\bf W}^H{\bf H}_b^H({\bm \Theta})  \right|}{\left|\sigma_e^2{\bf I}_{N_e}+{\bf H}_e({\bm \Theta}){\bf W}{\bf W}^H{\bf H}_e^H({\bm \Theta})  \right|} ,1 \right\},\\
\text{s.t.} \quad& \text{(\ref{cons1orig}), (\ref{cons2orig}), and (\ref{endorig}).}\label{reformualteobnew2}
\end{aligned}
\end{equation}
We observed that the objective function in problem (\ref{reformualteobnew2}) yields the maximum between the constant 1 and the fractional term $\frac{|\sigma_b^2{\bf I}_{N_b}+{\bf H}_b({\bm \Theta}){\bf W}{\bf W}^H{\bf H}_b^H({\bm \Theta})  |}{|\sigma_e^2{\bf I}_{N_e}+{\bf H}_e({\bm \Theta}){\bf W}{\bf W}^H{\bf H}_e^H({\bm \Theta})  |}$. Since only the fractional term depends on \( {\bf W} \) and \( {\bf \Theta} \), the problem can initially focus on maximizing this term. If the solution is less than 1, the secrecy rate becomes negative, rendering secure transmission unfeasible. Therefore, the problem can be reformulated by converting the maximization to a minimization and swapping the numerator and denominator in (\ref{reformualteobnew2}),
\begin{equation}
\begin{aligned}
\min _{{\bf W}, {\bm \Theta}} \quad &   \frac{\left|\sigma_e^2{\bf I}_{N_e}+{\bf H}_e({\bm \Theta}){\bf W}{\bf W}^H{\bf H}_e^H({\bm \Theta})  \right|}{\left|\sigma_b^2{\bf I}_{N_b}+{\bf H}_b({\bm \Theta}){\bf W}{\bf W}^H{\bf H}_b^H({\bm \Theta})  \right|},\\
\text{s.t.} \quad& \text{(\ref{cons1orig}), (\ref{cons2orig}), and (\ref{endorig}).}\label{reformualteobnew3}
\end{aligned}
\end{equation}
Since only the block-diagonal matrices of \( {\bf \Theta} \) are non-zero, to avoid directly handling the non-convex constraint in (\ref{cons2orig}), we optimize \( {\bm \Theta}_g \), \( \forall g \in \mathcal{G} \), i.e., the blockwise matrices of \( {\bf \Theta} \). Let ${\bm {\tilde \Theta}}=\{{\bf \Theta}_1,...,{\bf \Theta}_G\}$ denote the collection of \( {\bm \Theta}_g \). We equivalently rewrite the problem as follows,
\begin{subequations}
\begin{align}
\min _{{\bf W}, {\bm {\tilde \Theta}}} &\quad \frac{\left|{\bf I}_{N_e} +  {\bf \tilde H}_e({\bm {\tilde \Theta}}){\bf W}{\bf W}^H {\bf \tilde H}_e^H({\bm {\tilde \Theta}})\right|}{\left|{\bf I}_{N_b} +  {\bf \tilde H}_b({\bm {\tilde \Theta}}){\bf W}{\bf W}^H {\bf \tilde H}_b^H({\bm {\tilde \Theta}})\right|}, \\
\text{s.t.} &\quad \left\|{\bf W}\right\|^2 = 1, \label{refor31}\\
       &\quad {\bm \Theta}_g = {\bm \Theta}_g^T,{\bm \Theta}_g {\bm \Theta}_g^H = {\bf I}_{\tilde M}.\label{refor32}
\end{align}
\label{reformualteobnew3}%
\end{subequations}
where ${\bf \tilde H}_e({\bm {\tilde \Theta}})$ and ${\bf \tilde H}_b({\bm {\tilde \Theta}})$ are the effective channels, functions of the set of variables ${\bm {\tilde \Theta}} = \{{\bm \Theta}_1, \dots, {\bm \Theta}_G\}$, given by,
\begin{equation}
\begin{aligned}
    &{\bf \tilde H}_e({\bm {\tilde \Theta}}) = \sqrt{\frac{P}{\sigma_e^2}}\left(\sum_{g=1}^G {\bf \tilde H}_{ie}^g{\bf \Theta}_g{\bf \tilde H}_{ai}^g + {\bf H}_{ae}\right)\in \mathbb{C}^{N_e \times N_t},\\ 
    &{\bf \tilde H}_b({\bm {\tilde \Theta}}) = \sqrt{\frac{P}{\sigma_b^2}}\left(\sum_{g=1}^G {\bf \tilde H}_{ib}^g{\bf \Theta}_g{\bf \tilde H}_{ai}^g + {\bf H}_{ab}\right)\in \mathbb{C}^{N_b \times N_t},\\
    &{\bf \tilde H}_{ie}^g = {\bf H}_{ie}(:,(g-1)\tilde M+1:g\tilde M)\in \mathbb{C}^{N_e \times {\tilde M}}, \forall g \in \mathcal{G},\\
    &{\bf \tilde H}_{ib}^g = {\bf H}_{ib}(:,(g-1)\tilde M+1:g\tilde M)\in \mathbb{C}^{N_b \times {\tilde M}}, \forall g \in \mathcal{G},\\
    &{\bf \tilde H}_{ai}^g = {\bf H}_{ai}((g-1)\tilde M+1:g\tilde M, :)\in \mathbb{C}^{{\tilde M} \times N_t}, \forall g \in \mathcal{G}.
\end{aligned}
\end{equation}
Additionally, we have normalized the Frobenius norm magnitude of the beamformer \( {\bf W} \) to one by multiplying the channels with \( \sqrt{P} \). Furthermore, to decouple the constraint in (\ref{refor32}), we introduce a series of matrices \( {\bf \tilde \Psi}=\{ {\bf \Psi}_1, \dots, {\bf \Psi}_G \} \), where each \( {\bf \Psi}_g \in \mathbb{C}^{{\tilde M} \times {\tilde M}} \) for all \( g \in \mathcal{G} \), serving as copies of \( \{ {\bf \Theta}_1, \dots, {\bf \Theta}_G \} \), given as,
\begin{subequations}
\begin{align}
\min _{{\bf W}, {\bf \tilde \Theta},{\bf \tilde \Psi}} &\quad \frac{\left|{\bf I}_{N_e} +  {\bf \tilde H}_e({\bm {\tilde \Theta}}){\bf W}{\bf W}^H {\bf \tilde H}_e^H({\bm {\tilde \Theta}})\right|}{\left|{\bf I}_{N_b} +  {\bf \tilde H}_b({\bm {\tilde \Theta}}){\bf W}{\bf W}^H {\bf \tilde H}_b^H({\bm {\tilde \Theta}})\right|}, \label{objrefor4}\\
\text{s.t.} &\quad \left\|{\bf W}\right\|^2 = 1, \label{reofor41}\\
       &\quad {\bm \Theta}_g = {\bm \Theta}_g^T, \label{reofor42}\\ 
     &\quad {\bf \Psi}_g {\bf \Psi}_g^H = {\bf I}_{\tilde M},\label{reofor43}\\
      &\quad {\bf \Psi}_g  = {\bm \Theta}_g.\label{reofor44}
\end{align}
\label{reformualteobnew4}%
\end{subequations}
It can be observed that the constraints in (\ref{reofor41}), (\ref{reofor42}), and (\ref{reofor43}) naturally correspond to the complex sphere, the complex symmetric matrix, and the complex Stiefel manifolds within Riemannian space \cite{sato2021riemannian}. This motivates us to address problem (\ref{reformualteobnew4}) using manifold-based optimization methods \cite{sato2021riemannian}. However, the equivalent constraint in (\ref{reofor44}) prevents the direct application of these methods. To overcome this, we incorporate it as a penalty term within the objective function to penalize violations. Building on these insights, the P-PRCGD method is proposed in the following to efficiently solve problem (\ref{reformualteobnew4}).

\subsection{Proposed P-PRCGD Method}
In this section, we propose the P-PRCGD method to solve problem (\ref{reformualteobnew4}). Specifically, we begin by applying AL method \cite{boyd2004convex} to relax the equivalent constraint, incorporating it as a penalty term within the objective function. Next, we construct a smooth PRM that consolidates the information from the individual Riemannian manifolds associated with each constraint, transforming the problem into an unconstrained optimization problem within Riemannian space. Finally, the efficient PRCGD algorithm is developed to solve the problem within the Riemannian space.

\subsubsection{Riemannian-AL Transformation}
To handle the coupled equality constraint (\ref{reofor44}), the AL method is employed \cite{boyd2004convex}. Specifically, this approach transforms the constraints into penalty terms, balancing objective minimization with constraint satisfaction. The AL function for problem (\ref{reformualteobnew4}) is given by,
\begin{equation}
    \mathcal{L}({\bf W}, {\bf\tilde \Theta},{\bf \tilde\Psi}) = \left\{\begin{array}{l} f({\bf W}, {\bf \tilde\Theta},{\bf\tilde \Psi})\\ + \frac{1}{2\rho}\sum_{g=1}^G\left\| {\bf \Psi}_g-{\bf \Theta}_g  \right\|^2\\+\sum_{g=1}^G\Re\left\{\text{Tr}\left\{{\bf \Phi}_g^H({\bf \Psi}_g-{\bf \Theta}_g)\right\}\right\}\end{array} \right\}, \label{LRfuncu}
\end{equation}
where \(\rho > 0\) is the penalty parameter, and \({\bf \tilde\Phi}=\{{\bf \Phi}_1, \dots, {\bf \Phi}_G\}\) represents the collection of Lagrangian dual variables associated with (\ref{reofor44}), where \({\bf \Phi}_g \in \mathbb{C}^{{\tilde M} \times {\tilde M}}, \, \forall g \in \mathcal{G}\). The updates of \(\rho\) and \({\bf \Phi}_g\) will be discussed in Section III.B.(3). In the \(p\)-th iteration, the augmented Lagrangian problem with fixed penalty parameter \(\rho^p\) and dual variables \({\bf \tilde\Phi}^p\) is formulated as,
\begin{equation}
\begin{aligned}
  \min_{{\bf W}, {\bf \tilde\Theta},{\bf \tilde\Psi}} \quad&  \mathcal{L}({\bf W}, {\bf \tilde\Theta},{\bf \tilde\Psi}), \\
 \text{s.t.} \quad& \text{(\ref{reofor41}), (\ref{reofor42}), and (\ref{reofor43}).}\label{new1111}
  \end{aligned}
\end{equation}
Generally, the constraints in (\ref{reofor41}), (\ref{reofor42}), and (\ref{reofor43}) are non-convex, making them challenging to handle in Euclidean space. Manifolds, which can capture nonlinear relationships, provide a natural framework for incorporating these constraints. Specifically, constraints (\ref{reofor41}), (\ref{reofor42}), and (\ref{reofor43}) correspond to the complex sphere \( \mathcal{M}_{\mathbf{W}} \), the complex symmetric matrix \( \mathcal{M}_{{\mathbf{\Theta}}_g} \), and the complex Stiefel manifold \( \mathcal{M}_{{\mathbf{\Psi}}_g} \) Riemannian manifolds, respectively, as follows \cite{sato2021riemannian},
\begin{equation}
    \begin{aligned}
       & \mathcal{M}_{\bf W} = \left\{{\bf W} \in \mathbb{C}^{N_t \times N_s} \mid \|{\bf W}\|^2 = 1 \right\},\\
       & \mathcal{M}_{{\bf \Theta}_g} = \left\{{{\bf \Theta}_g} \in \mathbb{C}^{{\tilde M} \times {\tilde M}} \mid {\bf \Theta}_g = {\bf \Theta}_g^T \right\},\\
       & \mathcal{M}_{{\bf \Psi}_g} = \left\{{\bf \Psi}_g \in \mathbb{C}^{{\tilde M} \times {\tilde M}} \mid {\bf \Psi}_g {\bf \Psi}_g^H = {\bf I}_{\tilde M} \right\}. 
    \end{aligned}
\end{equation}
The Cartesian product of multiple individual Riemannian manifolds can form a PRM \cite{sato2021riemannian}. Based on this definition, the complex sphere Riemannian manifold \( \mathcal{M}_{\bf W} \), the \( G \) complex symmetric matrix Riemannian manifolds \( \mathcal{M}_{{\bf \Theta}_g}, \forall g \), and the \( G \) complex Stiefel Riemannian manifolds \( \mathcal{M}_{{\bf \Psi}_g}, \forall g \), can be combined to form a PRM \( \mathcal{M} \), which is given by,
\begin{equation}
    \mathcal{M} = \mathcal{M}_{\bf W} \times\prod_{g \in \mathcal{G}}\mathcal{M}_{{\bf \Theta}_g} \times \prod_{g \in \mathcal{G}} \mathcal{M}_{{\bf \Psi}_g}.   \label{productnow}
\end{equation}
Then, by combining \( \mathbf{W} \), \( (\mathbf{\Theta}_g^q)_{g=1}^G \), and \( (\mathbf{\Psi}_g^q)_{g=1}^G \) into,
\begin{equation}
{\bf \Upsilon} = \mathbf{W} \oplus \bigoplus_{g \in \mathcal{G}} \mathbf{\Theta}_g\oplus \bigoplus_{g \in \mathcal{G}} \mathbf{\Psi}_g ,\label{variablecomb}
\end{equation}
the problem (\ref{new1111}) is equivalently reformulated as,
\begin{equation}
  \min_{{\bf \Upsilon}\in\mathcal{M}} \quad  \mathcal{L}({\bf \Upsilon}),\label{nowopt111}
\end{equation}
which is an unconstrained problem over PRM \(\mathcal{M}\). 

Generally, optimizing over \( \mathcal{M} \) is challenging due to its non-linear, curved structure, which renders standard Euclidean methods inapplicable. Riemannian optimization overcomes this by operating on the tangent space \( \mathcal{T}_{\bf \Upsilon} \mathcal{M} \), a localized Euclidean approximation of the manifold at the current point \( \bf \Upsilon \) \cite{boumal2023introduction}. This approach allows for the adaptation of gradient-based methods by first computing a search direction (the Riemannian gradient) within this linear space and then mapping the result back onto the manifold via a retraction. A key property of a product manifold is that its tangent space is the direct sum of the tangent spaces of its constituent manifolds. This property provides the rationale for constructing \( \mathcal{T}_{\bf \Upsilon} \mathcal{M} \) as follows,
\begin{equation}
{\mathcal T}_{\bf \Upsilon}{\mathcal M} = {\mathcal T}_{\bf W}\mathcal{M}_{\bf W} \oplus\bigoplus_{g \in \mathcal{G}}{\mathcal T}_{{\bf \Theta}_g}\mathcal{M}_{{\bf \Theta}_g} \oplus \bigoplus_{g \in \mathcal{G}} {\mathcal T}_{{\bf \Psi}_g}\mathcal{M}_{{\bf \Psi}_g}, \label{tanspcae}
\end{equation}
where, 
\begin{equation}
    \begin{aligned}
       & {\mathcal T}_{\bf W}\mathcal{M}_{\bf W} = \{ {\bf \Xi}_{\bf W}\in\mathbb{C}^{N_t \times N_s} \mid \Re\{\text{Tr}\{{\bf \Xi}_{\bf W}^H{\bf W}\}\}= 0 \},\\
       & {\mathcal T}_{{\bf \Theta}_g}\mathcal{M}_{{\bf \Theta}_g} = \{ {\bf \Xi}_{{\bf \Theta}_g}\in\mathbb{C}^{{\tilde M} \times {\tilde M}} \mid {\bf \Xi}_{{\bf \Theta}_g}={\bf \Xi}_{{\bf \Theta}_g}^T \},\forall g,\\
       & {\mathcal T}_{{\bf \Psi}_g}\mathcal{M}_{{\bf \Psi}_g} = \{ {\bf \Xi}_{{\bf \Psi}_g}\in\mathbb{C}^{{\tilde M} \times {\tilde M}} \mid {\bf \Xi}_{{\bf \Psi}_g}^H{\bf \Psi}_g+{\bf \Psi}_g^H{\bf \Xi}_{{\bf \Psi}_g}={ 0} \},\forall g,
    \end{aligned}
\end{equation}
are the tangent spaces for individual manifolds \(\mathcal{M}_{\bf W}\), \(\mathcal{M}_{{\bf \Theta}_g} \), and \(\mathcal{M}_{{\bf \Psi}_g} \), respectively. Using \( \mathcal{M} \) and its tangent space \( \mathcal{T}_{\bf \Upsilon} \mathcal{M} \), we develop the PRCGD algorithm to solve problem (\ref{nowopt111}). The algorithm applies conjugate gradient descent (CGD) in \( \mathcal{T}_{\bf \Upsilon} \mathcal{M} \) and projects the solution back onto PRM. The details are provided below.

\begin{figure}[t]
  \begin{center}
  \includegraphics[width=2.5in]{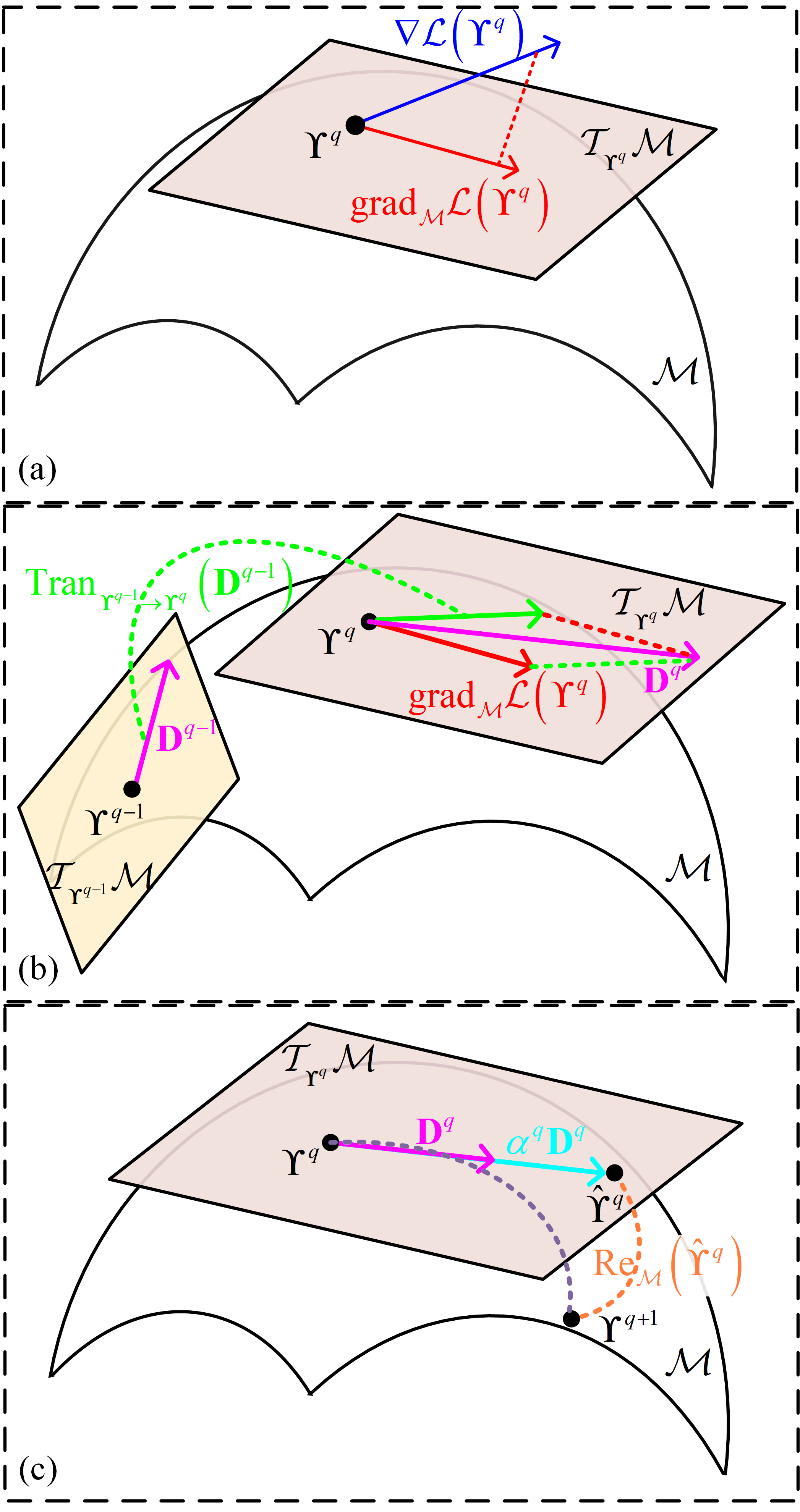}\\
  \caption{Illustration of (a) the calculation of the Riemannian gradient \( \text{grad}_{\mathcal{M}} \mathcal{L}({\bf \Upsilon}) \), (b) the derivation of the descent direction \( {\bf D}^{q} \), and (c) the update in the tangent space with a retraction \( \text{Re}_{\mathcal{M}}(\cdot) \) back onto the manifold.}\label{3}
  \end{center}
\end{figure}

\subsubsection{Proposed PRCGD algorithm}
Similar to the CGD algorithm in Euclidean spaces \cite{shewchuk1994introduction}, the PRCGD algorithm, as illustrated in Fig. \ref{3}, consists of three main steps when applied to \( \mathcal{M} \) at the \( q \)-th iteration. Firstly, the Riemannian gradient of the objective function is computed. Secondly, a descent direction is determined based on the Riemannian gradient. Finally, the update on the tangent space is completed, and a retraction is introduced to project the solution back onto the manifold. We repeat these procedures until convergence.

\textit{(a) Calculation of the Riemannian gradient:} To apply CGD to the Riemannian space, it is essential to calculate the steepest descent direction, which is determined by the Riemannian gradient \cite{boumal2023introduction}. As shown in Fig. \ref{3}(a), the Riemannian gradient is obtained by the orthogonal projection of the Euclidean gradient onto tangent space $\mathcal{T}_{{\bf \Upsilon}^q}\mathcal{M}$. In particular, the Riemannian gradient \( \text{grad}_{\mathcal{M}} \mathcal{L}({\bf \Upsilon}^q) \) of \( \mathcal{L}({\bf \Upsilon}^q) \) can be derived as follows,
\begin{equation}
\begin{aligned}
       \text{grad}_{\mathcal{M}} \mathcal{L}({\bf \Upsilon}^q)&= \left\{\begin{array}{l}  \text{grad}_{\mathcal{M}_{\bf W}}\mathcal{L}({\bf W}^q)\\\oplus\bigoplus_{g \in \mathcal{G}}\text{grad}_{\mathcal{M}_{{\bf \Theta}_g}}\mathcal{L}({\bf \Theta}_g^q) \\\oplus \bigoplus_{g \in \mathcal{G}} \text{grad}_{\mathcal{M}_{{\bf \Psi}_g}}\mathcal{L}({\bf \Psi}_g^q)\end{array} \right\}  \\& =\left\{\begin{array}{l}  \text{Proj}_{{\mathcal T}_{{\bf W}^q}\mathcal{M}_{\bf W} }(\nabla_{{\bf W}} \mathcal{L}({\bf W}^q))\\\oplus\bigoplus_{g \in \mathcal{G}}\text{Proj}_{{\mathcal T}_{{\bf \Theta}_g^q}\mathcal{M}_{{\bf \Theta}_g} }(\nabla_{{\bf \Theta}_g} \mathcal{L}({\bf \Theta}_g^q)) \\\oplus \bigoplus_{g \in \mathcal{G}} \text{Proj}_{{\mathcal T}_{{\bf \Psi}_g^q}\mathcal{M}_{{\bf \Psi}_g} }(\nabla_{{\bf \Psi}_g} \mathcal{L}({\bf \Psi}_g^q))\end{array} \right\},\label{calculReg}
\end{aligned}       
\end{equation}
where $\text{grad}_{\mathcal{M}_{\bf W}}\mathcal{L}({\bf W}^q)$, $\text{grad}_{\mathcal{M}_{{\bf \Theta}_g}}\mathcal{L}({\bf \Theta}_g^q)$ and $\text{grad}_{\mathcal{M}_{{\bf \Psi}_g}}\mathcal{L}({\bf \Psi}_g^q)$ are the Riemannian gradients of $\mathcal{L}({\bf W}^q)$, $\mathcal{L}({\bf \Theta}_g^q)$, and $\mathcal{L}({\bf \Psi}_g^q)$ on individual manifolds \(\mathcal{M}_{\bf W}\), \(\mathcal{M}_{{\bf \Theta}_g} \), and \(\mathcal{M}_{{\bf \Psi}_g} \), respectively, and,
\begin{equation}
\begin{aligned}
&    \text{Proj}_{{\mathcal T}_{{\bf W}^q}\mathcal{M}_{\bf W} }(\nabla_{{\bf W}} \mathcal{L}({\bf W}^q)) \\&=  \nabla_{{\bf W}} \mathcal{L}({\bf W}^q)-{\bf W}^q\Re\{\text{Tr}\{\nabla_{{\bf W}}^H \mathcal{L}({\bf W}^q){\bf W}^q\}\}, \\
 &   \text{Proj}_{{\mathcal T}_{{\bf \Theta}_g^q}\mathcal{M}_{{\bf \Theta}_g} }(\nabla_{{\bf \Theta}_g} \mathcal{L}({\bf \Theta}_g^q))= \frac{1}{2} ( \nabla_{{\bf \Theta}_g} \mathcal{L}({\bf \Theta}_g^q) +  \nabla_{{\bf \Theta}_g}^T \mathcal{L}({\bf \Theta}_g^q)  ),   \\
  &  \text{Proj}_{{\mathcal T}_{{\bf \Psi}_g^q}\mathcal{M}_{{\bf \Psi}_g} }(\nabla_{{\bf \Psi}_g} \mathcal{L}({\bf \Psi}_g^q)) \\&= \nabla_{{\bf \Psi}_g} \mathcal{L}({\bf \Psi}_g^q)-{\bf \Psi}_g^q \frac{1}{2}( \nabla_{{\bf \Psi}_g}^H \mathcal{L}({\bf \Psi}_g^q){\bf \Psi}_g^q+({\bf \Psi}_g^q)^H\nabla_{{\bf \Psi}_g} \mathcal{L}({\bf \Psi}_g^q)),  \label{calculReg2}
\end{aligned}
\end{equation}
are the orthogonal projections to project the Euclidean gradients (derived in the following \textit{Lemma \ref{lemma41}}) from the Euclidean space onto the tangent spaces ${\mathcal T}_{{\bf W}^q}\mathcal{M}_{\bf W}$, ${\mathcal T}_{{\bf \Theta}_g^q}\mathcal{M}_{{\bf \Theta}_g}$, and ${\mathcal T}_{{\bf \Psi}_g^q}\mathcal{M}_{{\bf \Psi}_g}$, respectively.  

\begin{lemma} \label{lemma41}
The Euclidean gradients \( \nabla_{{\bf W}} \mathcal{L}({\bf W}) \), \( \nabla_{{\bf \Theta}_g} \mathcal{L}({\bf \Theta}_g)  \), and \( \nabla_{{\bf \Psi}_g} \mathcal{L}({\bf \Psi}_g)  \), with respect to variables \( {\bf W} \), \( {\bf \Theta}_g \), and \( {\bf \Psi}_g \) are provided in equations (\ref{EDW}) , (\ref{EDTHETA}), and (\ref{EDPSI}), respectively.
\end{lemma}

\textit{Proof}: See Appendix \ref{appendixA}. $\hfill\blacksquare$

\textit{(b) Derivation of the descent direction:} Traditionally, the steepest descent direction, i.e., the negative Riemannian gradient, is used as the descent direction in gradient-based manifold optimization methods \cite{boumal2023introduction}. However, its practical convergence is often hindered by a relatively slow rate and susceptibility to oscillations around local minima. To overcome this limitation, a more efficient approach is to use the CGD direction. As shown in Fig. \ref{3}(b), the CGD direction is given by,
\begin{equation}
    {\bf D}^{q} = -\text{grad}_{\mathcal{M}} \mathcal{L}({\bf \Upsilon}^q)+\beta^q \text{Tran}_{{{\bf \Upsilon}^{q-1}} \to {{\bf \Upsilon}^q}}({\bf D}^{q-1}),\label{desRe}
\end{equation}
where $\text{grad}_{\mathcal{M}} \mathcal{L}({\bf \Upsilon}^q)$ is the Riemannian gradient derived in (\ref{calculReg}), $\beta$ is the Fletcher-Reeves parameter, given as \cite{shewchuk1994introduction},
\begin{equation}
    \beta^q = \frac{|| \text{grad}_{\mathcal{M}} \mathcal{L}({\bf \Upsilon}^q)||^2}{|| \text{grad}_{\mathcal{M}} \mathcal{L}({\bf \Upsilon}^{q-1})||^2},\label{conjgateparemae}
\end{equation}
and $\text{Tran}_{{{\bf \Upsilon}^{q-1}} \to {{\bf \Upsilon}^q}}({\bf D}^{q-1})$ is the vector transport operation to relocate the previous descent direction ${\bf D}^{q-1}\in\mathcal{M}$ from the tangent space $\mathcal{T}_{{\bf \Upsilon}^{q-1}}\mathcal{M}$ to the tangent space $\mathcal{T}_{{\bf \Upsilon}^{q}}\mathcal{M}$, given as,
\begin{equation}
    \text{Tran}_{{{\bf \Upsilon}^{q-1}} \to {{\bf \Upsilon}^q}}({\bf D}^{q-1})=\left\{\begin{array}{l}  \text{Tran}_{{{\bf W}^{q-1}} \to {{\bf W}^q}}({\bf D}_{\bf W}^{q-1})\\\oplus\bigoplus_{g \in \mathcal{G}}\text{Tran}_{{{\bf \Upsilon}^{q-1}} \to {{\bf \Upsilon}^q}}({\bf D}_{{\bf \Theta}_g}^{q-1}) \\\oplus \bigoplus_{g \in \mathcal{G}} \text{Tran}_{{{\bf \Upsilon}^{q-1}} \to {{\bf \Upsilon}^q}}({\bf D}_{{\bf \Psi}_g}^{q-1})\end{array} \right\},
\end{equation}
where,
\begin{equation}
    \begin{aligned}
       & \text{Tran}_{{{\bf W}^{q-1}} \to {{\bf W}^q}}({\bf D}_{\bf W}^{q-1})={\bf D}_{\bf W}^{q-1}-{\bf W}^q\Re\{\text{Tr}\{{\bf D}_{\bf W}^{q-1}{\bf W}^q\}\},\\
    &\text{Tran}_{{{\bf \Upsilon}^{q-1}} \to {{\bf \Upsilon}^q}}({\bf D}_{{\bf \Theta}_g}^{q-1})=\frac{1}{2} ( {\bf D}_{{\bf \Theta}_g}^{q-1} +  ({\bf D}_{{\bf \Theta}_g}^{q-1})^T  ),\\
    & \text{Tran}_{{{\bf \Upsilon}^{q-1}} \to {{\bf \Upsilon}^q}}({\bf D}_{{\bf \Psi}_g}^{q-1})\\&={\bf D}_{{\bf \Psi}_g}^{q-1}-{\bf \Psi}_g^q \frac{1}{2}( ({\bf D}_{{\bf \Psi}_g}^{q-1})^H{\bf \Psi}_g^q+({\bf \Psi}_g^q)^H{\bf D}_{{\bf \Psi}_g}^{q-1}),\label{usetofirgue2}
    \end{aligned}
\end{equation}
are the vector transports for individual manifolds \(\mathcal{M}_{\bf W}\), \(\mathcal{M}_{{\bf \Theta}_g} \), and \(\mathcal{M}_{{\bf \Psi}_g} \), respectively. Specifically, transporting the previous descent direction \( {\bf D}^{q-1} \) between tangent spaces is essential because, due to the non-linear nature of the manifold, linear operations are only applicable within the tangent space. The Riemannian gradient \( \text{grad}_{\mathcal{M}} \mathcal{L}({\bf \Upsilon}) \) and the previous descent direction \( {\bf D}^{q-1} \) belong to different tangent spaces, \( \mathcal{T}_{{\bf \Upsilon}^{q}} \mathcal{M} \) and \( \mathcal{T}_{{\bf \Upsilon}^{q-1}} \mathcal{M} \), respectively. Therefore, it is necessary to transport \( {\bf D}^{q-1} \) from the previous tangent space to the current tangent space so that these vectors can be combined.

\textit{(c) Update in the tangent space and retraction:} With the descent direction derived in (\ref{desRe}), as shown in Fig. \ref{3}(c), the update in the tangent space \( \mathcal{T}_{{\bf \Upsilon}^{q}} \mathcal{M} \) is given by,
\begin{equation}
    {  \bf \hat \Upsilon}^{q} = {\bf \Upsilon}^{q} + \alpha^q {\bf D}^{q},\label{Updateupsi}
\end{equation}
where \( \alpha^q \) is the step size, and its calculation will be described later. Since PRM \( \mathcal{M} \) is non-linear, i.e., \( (\alpha^q {\bf D}^q + {\bf \Upsilon}^q) \notin \mathcal{M} \), even though \( \alpha^q {\bf D}^q, {\bf \Upsilon}^q \in \mathcal{M} \). As a result, the updated point \( {\bf \hat{\Upsilon}}^{q} \) may fall outside of \( \mathcal{M} \). In this case, a retraction is introduced to map the updated point from \( \mathcal{T}_{{\bf \Upsilon}^q} \mathcal{M} \) back into \( \mathcal{M} \), as given by,
\begin{equation}
   {\bf \Upsilon}^{q+1}= \text{Re}_{\mathcal{M}}({  \bf \hat \Upsilon}^{q})=\left\{\begin{array}{l}\text{Re}_{\mathcal{M}_{\bf W}}({\bf \hat W}^{q})\\\oplus\bigoplus_{g \in \mathcal{G}}\text{Re}_{\mathcal{M}_{{\bf  \Theta}_g}}({\bf \hat\Theta}_g^q)\\\oplus\bigoplus_{g \in \mathcal{G}}\text{Re}_{\mathcal{M}_{{\bf \Psi}_g}}({\bf \hat \Psi}_g^q)\end{array} \right\}, \label{useforretrac}
\end{equation}
where \( \text{Re}_{\mathcal{M}_{\bf W}}({\bf \hat W}^{q}) \), \( \text{Re}_{\mathcal{M}_{{\bf \Theta}_g}}({\bf \hat\Theta}_g^q) \), and \( \text{Re}_{\mathcal{M}_{{\bf \Psi}_g}}({\bf \hat \Psi}_g^q) \) are retractions that map the variables \( {\bf \hat W}^{q} \), \( {\bf \hat \Theta}_g^q \), and \( {\bf \hat \Psi}_g^q \) back onto their respective manifolds \( \mathcal{M}_{\bf W} \), \( \mathcal{M}_{{\bf \Theta}_g} \), and \( \mathcal{M}_{{\bf \Psi}_g} \), respectively, as given by,
\begin{equation}
    \begin{aligned}
        &\text{Re}_{\mathcal{M}_{\bf W}}({\bf \hat W}^{q})={\bf \hat W}^{q}\oslash{\|{\bf \hat W}^{q}\|^2},\\&\text{Re}_{\mathcal{M}_{{\bf \Theta}_g}}({\bf \hat \Theta}_g^{q})={\bf \hat \Theta}_g^{q},\\&\text{Re}_{\mathcal{M}_{{\bf \Psi}_g}}({\bf \hat \Psi}_g^{q})=\text{qf}({\bf \hat \Psi}_g^{q}),\label{retracnow}
    \end{aligned}
\end{equation}
where $\text{qf}({\bf \hat \Psi}_g^{q})$ is the $Q$ factor of the $QR$ factorization of ${\bf \hat \Psi}_g^{q}$.

To determine the step size, we employ the backtracking line search strategy, which dynamically adjusts the step size during updates to ensure that each iteration satisfies the Riemannian Wolfe conditions \cite{ring2012optimization}. This adaptive approach maintains a sufficient decrease in the objective function value and ensures that the step size does not exceed a predefined bound in each iteration. In particular, the chosen
step size \( \alpha^q \) has to satisfy,
\begin{subequations}
\begin{align}
    &\mathcal{L}({\bf \Upsilon}^{q+1}) -\mathcal{L}({\bf \Upsilon}^{q})\le \sigma_1\alpha^q \langle\text{grad}_{\mathcal{M}} \mathcal{L}({\bf \Upsilon}^q), {\bf D}^{q} \rangle,\label{refstepsize1}\\  
    &|\langle\text{grad}_{\mathcal{M}} \mathcal{L}({\bf \Upsilon}^{q+1}),\text{Tran}_{q\to q+1}({\bf D}^{q})  \rangle|\notag\\&\quad\quad\quad\quad\quad\quad\quad\quad\le\sigma_2|\langle\text{grad}_{\mathcal{M}} \mathcal{L}({\bf \Upsilon}^{q}),{\bf D}^{q}  \rangle|,\label{refstepsize}
\end{align}   
\label{RWCONdi}%
\end{subequations}
where $\sigma_1,\sigma_2\in(0,1)$.

In summary, the proposed PRCGD algorithm for solving (\ref{new1111}) is summarized in Algorithm \ref{alg:1}.

\begin{algorithm}
	\floatname{algorithm}{Algorithm}
	\renewcommand{\algorithmicrequire}{\textbf{Input:}}
	\renewcommand{\algorithmicensure}{\textbf{Output:}}
	\caption{: The PRCGD algorithm to the problem (\ref{new1111}).}
	\label{alg:1}
	\begin{algorithmic}[1]
		\REQUIRE 
                ${\bf W}^p, {\bf \tilde \Theta}^p,{\bf \tilde \Psi}^p $.\\
            \STATE Initialize $q = 0$, ${\bf W}^q={\bf W}^p$, ${\bf \tilde \Theta}^q={\bf \tilde \Theta}^p$, and ${\bf \tilde \Psi}^q={\bf \tilde \Psi}^p$;
            \STATE Combine ${\bf W}^q$, ${\bf \tilde \Theta}^q$, and ${\bf \tilde \Psi}^q$ into ${\bf \Upsilon}^q$ by (\ref{variablecomb});
            \STATE {\textbf {While $\|\text{grad}_{\mathcal{M}} \mathcal{L}({\bf \Upsilon}^q)\|\ge\varepsilon^p$ do}}
            \STATE \quad Update $\text{grad}_{\mathcal{M}} \mathcal{L}({\bf \Upsilon}^q)$ by (\ref{calculReg}) and (\ref{calculReg2});
            \STATE \quad Update ${\bf D}^{q}$ by (\ref{desRe})-(\ref{usetofirgue2});
            \STATE \quad Update $\alpha^q$ by (\ref{refstepsize});
            \STATE \quad Update ${\bf \Upsilon}^{q+1}$ by (\ref{Updateupsi})-(\ref{retracnow});
            \STATE \quad $q \leftarrow q+1$;
            \STATE {\textbf {End while}}
            \STATE Decompose ${\bf \Upsilon}^{q}$ into ${\bf W}^{q}$, ${\bf \tilde \Theta}^{q}$, and ${\bf \tilde \Psi}^{q}$;
        \ENSURE
                ${\bf W}^{p+1}={\bf W}^{q}$, ${\bf \tilde \Theta}^{p+1}={\bf \tilde \Theta}^{q}$, ${\bf \tilde \Psi}^{p+1}={\bf \tilde \Psi}^{q}$.\\
	\end{algorithmic}%
\end{algorithm}

\subsubsection{Parameter Update}
To update the Lagrangian dual variable ${\bf \tilde\Phi} = \{ {\bf \Phi}_1, \dots, {\bf \Phi}_G \}$ and the penalty parameter $\rho > 0$ within the P-PRCGD method, we proceed as follows. In the $p$-th iteration, if the maximum violation of the equality constraint, defined as $\eta^p = \max \{ \| {\bf \Psi}^p_g - {\bf \Theta}^p_g \|_{\infty}, \forall g \}$, is significantly large, $\rho$ is decreased to enforce a stronger penalty on constraint violations, i.e.,
\begin{equation}
\rho^{p+1} = \gamma_1 \cdot \rho^p,
\end{equation}
where \( \gamma_1 \in (0, 1) \). Conversely, when the violation is relatively small, the dual variable \( {\bf \tilde\Phi} \) is updated to improve convergence stability and ensure a more robust optimization process, as follows,
\begin{equation}
{\bf \Phi}_g^{p+1} = {\bf \Phi}_g^p + \frac{1}{\rho} \left( {\bf \Theta}_g^p - {\bf \Psi}_g^p \right), \forall g.
\end{equation}
Since the constraint violations \( \eta^p \) vanish as \( \rho \) approaches infinity, one of the termination conditions for the P-PRCGD method is \( \eta^p < \eta_{\min} \to0\), where \( \eta_{\min} \) is a small prescribed constant.

The threshold parameter \( \varepsilon^p \) in Algorithm \ref{alg:1} is used to limit the Riemannian gradient Frobenius norm of the PRCGD algorithm to satisfy \( \| \text{grad}_{\mathcal{M}} \mathcal{L}({\bf \Upsilon}^q) \| < \varepsilon^p \to \varepsilon_{\min} = 0 \), where \( \varepsilon_{\min} \) is a small prescribed constant. As iterations progress, more accurate solutions are required, which necessitates a reduction in \( \varepsilon^p \), i.e.,
\begin{equation}
\varepsilon^{p+1} = \gamma_2 \cdot \varepsilon^p,
\end{equation}
where \( \gamma_2 \in (0, 1) \). Thus, another termination condition for the P-PRCGD method is \( \varepsilon^p \leq \varepsilon_{\min} \). With the above discussions, the procedure for the proposed P-PRCGD method is straightforward and summarized in Algorithm \ref{alg:2}.

\textit{\textbf{Remark 5:}} It is important to clarify that our proposed P-PRCGD algorithm is specifically tailored for the BD-RIS architecture with group size ${\tilde M} > 1$ and does not efficiently simplify to the D-RIS case, which corresponds to ${\tilde M} = 1$. The core of our framework, particularly the AL penalty method, is designed to resolve the coupled symmetric and stiefel manifold constraints unique to BD-RIS. In the D-RIS scenario, these complex constraints vanish, reducing to simple unit-modulus constraints on a complex circle manifold \cite{xiong2025constant}. Consequently, the entire rationale for our algorithm's double-loop penalty structure becomes obsolete. Therefore, our algorithm should be viewed as a specialized solution for the unique structural complexities of BD-RIS, complementary to existing methods designed for D-RIS.

\begin{algorithm}
	\floatname{algorithm}{Algorithm}
	\renewcommand{\algorithmicrequire}{\textbf{Input:}}
	\renewcommand{\algorithmicensure}{\textbf{Output:}}
	\caption{: The P-PRCGD method to the problem (\ref{reformualteobnew4}).}
	\label{alg:2}
	\begin{algorithmic}[1]
		\REQUIRE 
                ${\bf W}^0, {\bf \tilde \Theta}^0,{\bf \tilde \Psi}^0,{\bf \tilde\Phi}^0,\rho^0,\eta^{0},\varepsilon^{0},\gamma_1,\gamma_2,\gamma_3,\eta_{\min},\varepsilon_{\min}$.\\
            \STATE Initialize outer iteration number $p=0$;\\
            \STATE {\textbf {While $\eta^{p}\ge\eta_{\min}$ or $\varepsilon^{p}\ge\varepsilon_{\min}$ do}}
            \STATE \quad Update ${\bf W}^{p+1}$, ${\bf \tilde \Theta}^{p+1}$, and ${\bf \tilde \Psi}^{p+1}$ by \textbf{Algorithm \ref{alg:1}};
            \STATE \quad Update $\eta^{p+1} = \max \{ \| {\bf \Psi}^{p+1}_g - {\bf \Theta}^{p+1}_g \|_{\infty}, \forall g \}$;
            \STATE \quad \textbf{If} $\eta^{p+1}\ge\gamma_3\cdot\eta^{p}$;
            \STATE \quad\quad Update $\rho^{p+1} = \gamma_1 \cdot \rho^p$;
            \STATE \quad \textbf{Else}
            \STATE \quad\quad Update ${\bf \Phi}_g^{p+1} = {\bf \Phi}_g^p + \frac{1}{\rho} \left( {\bf \Theta}_g^p - {\bf \Psi}_g^p \right), \forall g\in \mathcal{G}$ ;
            \STATE \quad \textbf{End}
            \STATE \quad $\varepsilon^{p+1} = \gamma_2 \cdot \varepsilon^p$;
            \STATE \quad $p \leftarrow p+1$;
            \STATE {\textbf {End while}}
        \ENSURE ${\bf W}^{p}$, ${\bf \tilde \Theta}^{p}$, ${\bf \tilde \Psi}^{p}$.
	\end{algorithmic}%
\end{algorithm}

\subsection{Computational Complexity and Convergence Analysis of P-PRCGD}
In this section, we analyze the computational complexity and the convergence of the P-PRCGD method. The details are given in the following.

\subsubsection{Computational complexity analysis of P-PRCGD method}
The computational complexity of the proposed P-PRCGD method primarily stems from Algorithm \ref{alg:1}, with the main cost arising from the computation of the Riemannian gradient $\text{grad}_{\mathcal{M}} \mathcal{L}({\bf \Upsilon})$ in (\ref{calculReg}). Since $N_e, N_b, N_s \ll M, N_t$ in practice, we focus on the dominant parameters $M$ and $N_t$. The computation of the Riemannian gradient $\text{grad}_{\mathcal{M}} \mathcal{L}({\bf \Upsilon})$ involves evaluating the individual gradients: $\text{grad}_{\mathcal{M}_{\bf W}} \mathcal{L}({\bf W})$, $G$ instances of $\text{grad}_{\mathcal{M}_{{\bf \Theta}_g}} \mathcal{L}({\bf \Theta}_g)$, and $G$ instances of $\text{grad}_{\mathcal{M}_{{\bf \Psi}_g}} \mathcal{L}({\bf \Psi}_g)$. Specifically, the complexity of computing $\text{grad}_{\mathcal{M}_{\bf W}} \mathcal{L}({\bf W})$ is $\mathcal{O}({\tilde M}^2 + {\tilde M} N_t + 3 N_t)$; the complexity for the $G$ instances of $\text{grad}_{\mathcal{M}_{{\bf \Theta}_g}} \mathcal{L}({\bf \Theta}_g)$ is $\mathcal{O}({\tilde M}^2N_t)$; and the complexity for the $G$ instances of $\text{grad}_{\mathcal{M}_{{\bf \Psi}_g}} \mathcal{L}({\bf \Psi}_g)$ is $\mathcal{O}({\tilde M})$. Therefore, the overall complexity of computing $\text{grad}_{\mathcal{M}} \mathcal{L}({\bf \Upsilon})$ is approximately $\mathcal{O}({\tilde M}^2 N_t)$ per inner iteration. Assuming $J$ inner-loop iterations in Algorithm \ref{alg:1} and $I$ outer-loop iterations in Algorithm \ref{alg:2}, the total computational complexity of the P-PRCGD method is approximately $\mathcal{O}(IJ{\tilde M}^2N_t)$. The detailed component-wise and total complexities are summarized in Table \ref{tab:complexity_professional}.

\begin{table}[h!]
  \centering
  \caption{Computational complexities involved in the P-PRCGD method.}
  \label{tab:complexity_professional}
  \begin{tabular}{l l}
    \toprule 
    \textbf{Tasks} & \textbf{Complexities} \\
    \midrule 
    Compute $\text{grad}_{\mathcal{M}_{\bf W}} \mathcal{L}({\bf W})$ & $\mathcal{O}({\tilde M}^2 + {\tilde M} N_t + 3 N_t)$  \\
    Compute $\text{grad}_{\mathcal{M}_{{\bf \Theta}_g}} \mathcal{L}({\bf \Theta}_g)$ & $\mathcal{O}({\tilde M}^2N_t)$ \\
    Compute $\text{grad}_{\mathcal{M}_{{\bf \Psi}_g}} \mathcal{L}({\bf \Psi}_g)$ & $\mathcal{O}({\tilde M})$ \\
    \midrule 
    Overall algorithm & $\mathcal{O}(IJ{\tilde M}^2 N_t)$ \\
    \bottomrule 
  \end{tabular}
\end{table}

\subsubsection{Convergence analysis of P-PRCGD method}
To establish the proof, we first show that Algorithm \ref{alg:1} is able to converge to a stationary point in \textit{Theorem \ref{theorem1}}. Based on this theorem, we then show that every limit point generated by Algorithm \ref{alg:2} is a stationary point of the original problem in \textit{Theorem \ref{theorem2}}.

\begin{theorem} \label{theorem1}
 \textit{(Convergence of PRCGD algorithm)} Let \( \{{\bf \Upsilon}^q\} \) denote the sequence generated by Algorithm \ref{alg:1}, where the initial descent direction is \( {\bf D}^1 = -\text{grad}_{\mathcal{M}} \mathcal{L}({\bf \Upsilon}^1) \), and the parameter \( \sigma_2 \) satisfies \( 0 < \sigma_2 < \frac{1}{2} \). Assume that the Riemannian gradient of \( \mathcal{L}({\bf \Upsilon}) \) is Lipschitz continuous, i.e.,
\begin{equation}
\|\text{grad}_{\mathcal{M}} \mathcal{L}({\bf \Upsilon}^q) - \text{grad}_{\mathcal{M}} \mathcal{L}({\bf \Upsilon}^{q-1})\| \leq K \|{\bf \Upsilon}^q - {\bf \Upsilon}^{q-1}\|, \forall q, \label{Lipschitzc}
\end{equation}
for some constant \( K > 0 \). Then,
\begin{equation}
\lim_{q \to \infty} \inf \|\text{grad}_{\mathcal{M}} \mathcal{L}({\bf \Upsilon}^q)\| = 0,
\end{equation}
which ensures that Algorithm \ref{alg:1} converges to a stationary point of problem (\ref{new1111}).
\end{theorem}

\textit{Proof}: See Appendix \ref{appendixB}. $\hfill\blacksquare$

\begin{theorem} \label{theorem2}
 \textit{(Convergence of P-PRCGD method)} Let \( \{{\bf \Upsilon}^p\} \) denote the sequence generated by Algorithm \ref{alg:2}, with the inner loop updated by Algorithm \ref{alg:1}. Suppose that \( {\bf \Upsilon}^* \) is a limit point of the sequence \( \{{\bf \Upsilon}^p\} \), where the LICQ condition is satisfied. Then, \( {\bf \Upsilon}^* \) is a stationary point of the problem (\ref{reformualteobnew4}).
\end{theorem}

\textit{Proof}: See Appendix \ref{appendixC}. $\hfill\blacksquare$

\textit{\textbf{Remark 6:}} The theoretical convergence to a stationary point directly supports the enhanced SR achieved by our method. This connection manifests in two key aspects. Firstly, as a descent-based algorithm, our method ensures that the SR is monotonically non-decreasing across iterations. More importantly, it guarantees convergence to a stationary point, which corresponds to a high-quality local optimum where no further performance gain is possible. This property fundamentally distinguishes our approach from conventional methods such as alternating optimization (AO), which, although monotonic, lack strong convergence guarantees and may stall at suboptimal points. This is further demonstrated in the numerical results (see Section V), where the proposed method consistently achieves higher SR than AO.

\section{Extensions to Imperfect CSI Scenario}
Although various techniques can be employed to estimate the channels involving BD-RIS, CEEs are generally unavoidable in practice \cite{mukherjee2012detecting}. This makes perfect CSI difficult to obtain and leads to a mismatch between the analytical design and the actual deployment. Therefore, it is important to account for CEEs in the optimization design to achieve more robust system performance \cite{yoo2006capacity}.

\subsection{Problem Formulation to Imperfect CSI Scenario}
To extend the design to the imperfect CSI scenario and examine the impact of CEEs, let $\Delta {\bf H}_{ai}$, $\Delta {\bf H}_{ir}$, and $\Delta {\bf H}_{ar}$ denote the CEEs corresponding to ${\bf H}_{ai}$, ${\bf H}_{ir}$, and ${\bf H}_{ar}$, respectively, where $r \in {b, e}$ and \( r \) indexes \( e \) (Eve) or \( b \) (Bob). The actual channel can then be modeled as \cite{yoo2006capacity},
\begin{equation}
    \left\{
\begin{aligned}
&{\bf \hat H}_{ai} = {\bf H}_{ai} + \Delta {\bf H}_{ai}, \\
& {\bf \hat H}_{ir} = {\bf H}_{ir} + \Delta {\bf H}_{ir}, \\
&{\bf \hat H}_{ar} = {\bf H}_{ar} + \Delta{\bf H}_{ar},
\end{aligned}
\right.
\end{equation}
Under this model, the received signals at Bob and Eve, considering imperfect CSI, can be reformulated as,
\begin{equation}
\begin{aligned}
    &{\mathbf{\hat y}}_r= ( {\bf \hat H}_{ir} {\bf{\Theta}} {\bf \hat H}_{ai} + {\bf \hat H}_{ar} ) \mathbf{W}\mathbf{s}+ {\bf{n}}_b,\\
   &={\bf H}_r({\bm \Theta})\mathbf{W}\mathbf{s}+\underbrace{\left[\begin{array}{l}  \Delta{\bf H}_{ir}{\bf{\Theta}}{\bf H}_{ai}\\+\Delta {\bf H}_{ir}{\bf{\Theta}}\Delta {\bf H}_{ai}\\+{\bf H}_{ir}{\bf{\Theta}}\Delta {\bf H}_{ai}  + \Delta{\bf H}_{ar}\end{array} \right]\mathbf{W}\mathbf{s}+ {\bf{n}}_b}_{\text{Interference and noise: } \mathbf{N}_r}. \label{recieveimCSI}
\end{aligned}   
\end{equation}
According to the information-theoretic analysis in \cite{hassibi2003much}, CEEs resulting from minimum mean square error (MMSE) channel estimation follow a circularly symmetric complex Gaussian (CSCG) distribution. Since MMSE estimation also applies to RIS-related channels \cite{zhao2021exploiting}, modeling CEEs as CSCG-distributed is well justified. To compactly characterize CEEs in the BD-RIS-assisted MIMO system, we adopt the classical Gaussian-Kronecker model from \cite{wu2021intelligent}, given by,
\begin{equation}
        \left\{
\begin{aligned}
&\Delta {\bf H}_{ai}  \sim \mathcal{CN}(0,{\bf A}_{ai}\otimes{\bf B}_{ai}), \\
& \Delta {\bf H}_{ir}  \sim \mathcal{CN}(0,{\bf A}_{ir}\otimes{\bf B}_{ir}), \\
&\Delta {\bf H}_{ar}  \sim \mathcal{CN}(0,{\bf A}_{ar}\otimes{\bf B}_{ar}),
\end{aligned}
\right. \label{usenow1now}
\end{equation}
where ${\bf A}_{ai}$, ${\bf A}_{ir}$, and ${\bf A}_{ar}$ denote the receive-side covariance matrices, while ${\bf B}_{ai}$, ${\bf B}_{ir}$, and ${\bf B}_{ar}$ represent the transmit-side covariance matrices. For insight, consider a simplified case where the entries of the estimation errors $\Delta {\bf H}_{ai}$, $\Delta {\bf H}_{ir}$, and $\Delta {\bf H}_{ar}$ are i.i.d. zero-mean CSCG variables with variances $\sigma_{ai}^2$, $\sigma_{ir}^2$, and $\sigma_{ar}^2$, respectively. The corresponding spatial correlation matrices are given by \cite{zeng2022joint},
\begin{equation}
        \left\{
\begin{aligned}
&{\bf A}_{ai}={\bf I},\quad{\bf B}_{ai}=\sigma_{ai}^2\cdot{\bf I}, \\
& {\bf A}_{ir}={\bf I},\quad{\bf B}_{ir}=\sigma_{ir}^2\cdot{\bf I}, \\
&{\bf A}_{ar}={\bf I},\quad{\bf B}_{ar}=\sigma_{ar}^2\cdot{\bf I},
\end{aligned}
\right. \label{simpleCSGs}
\end{equation}
Based on (\ref{recieveimCSI})~(\ref{simpleCSGs}), the channel capacities of Bob and Eve by considering CEEs can be formulated as \cite{shi2011iteratively},
\begin{equation}
    {R_r({\bf W},{\bm \Theta})}= \log \left|{\bf I}_{N_r}+{\bf H}_r({\bm \Theta}){\bf W}{\bf W}^H{\bf H}_r^H({\bm \Theta}){\bf J}_r^{-1}  \right|, \label{imcSIcp}
\end{equation}
where ${\bf J}_r = \mathbb{E}\{{\bf N}_r {\bf N}_r^H\}$ denotes the covariance matrix of the interference and noise. By adopting the proposed statistical Gaussian CEE model and substituting ${\bf N}_r$ from (\ref{recieveimCSI}) into ${\bf J}_r$, we obtain, 
\begin{equation}
    \begin{aligned}
        {\bf J}_r =& \mathbb{E}\{\Delta {\bf H}_{ir}{\bf{\Theta}}\Delta {\bf H}_{ai}{\bf W}{\bf W}^H(\Delta {\bf H}_{ir}{\bf{\Theta}}\Delta {\bf H}_{ai})^H\}
        \\&+\mathbb{E}\{\Delta{\bf H}_{ir}{\bf{\Theta}}{\bf H}_{ai}{\bf W}{\bf W}^H(\Delta{\bf H}_{ir}{\bf{\Theta}}{\bf H}_{ai})^H\}
        \\&+\mathbb{E}\{{\bf H}_{ir}{\bf{\Theta}}\Delta {\bf H}_{ai}{\bf W}{\bf W}^H({\bf H}_{ir}{\bf{\Theta}}\Delta {\bf H}_{ai})^H\}
        \\&+\mathbb{E}\{\Delta{\bf H}_{ar}{\bf W}{\bf W}^H\Delta{\bf H}_{ar}^H\}+\sigma_r^2{\bf I},
    \end{aligned}
\end{equation}
According to \cite{rong2011robust}, for $\Delta {\bf H}  \sim \mathcal{CN}(0,{\bf A}\otimes{\bf B})$, there is $\mathbb{E}\{\Delta {\bf H}{\bf X}\Delta {\bf H}^H\}=\text{Tr}({\bf X}{\bf A}^T){\bf B}$. Based on this lemma, ${\bf J}_r$ can be calculated as,
\begin{equation}
        \begin{aligned}
        {\bf J}_r =&\text{Tr}\{{\bf{\Theta}}\text{Tr}\{{\bf W}{\bf W}^H{\bf A}_{ai}^T\}{\bf B}_{ai}{\bf{\Theta}}^H{\bf A}_{ir}^T\}{\bf B}_{ir}
        \\& + \text{Tr}\{{\bf{\Theta}}{\bf H}_{ai}{\bf W}{\bf W}^H{\bf H}_{ai}^H{\bf{\Theta}}^H{\bf A}_{ir}^T\}{\bf B}_{ir}
        \\&+ \text{Tr}\{{\bf W}{\bf W}^H{\bf A}_{ai}^T\}{\bf H}_{ir}^H{\bf{\Theta}}{\bf B}_{ai}{\bf{\Theta}}^H{\bf H}_{ir} 
        \\&+\text{Tr}({\bf W}{\bf W}^H{\bf A}_{ar}^T){\bf B}_{ar}+\sigma_r^2{\bf I}.\label{firstversion}
    \end{aligned}
\end{equation}
Since the BD-RIS is subject to the structural constraint ${\bf \Theta}{\bf \Theta}^H = {\bf I}$, and based on the definition of spatial correlation matrices in (\ref{simpleCSGs}), ${\bf J}_r$ in (\ref{firstversion}) can be further simplified as,
\begin{equation}
        \begin{aligned}
        {\bf J}_r =&\sigma_{ai}^2\sigma_{ir}^2M\text{Tr}({\bf W}{\bf W}^H){\bf I}
        + \sigma_{ir}^2\text{Tr}({\bf H}_{ai}{\bf W}{\bf W}^H{\bf H}_{ai}^H){\bf I}
        \\&+ \sigma_{ai}^2\text{Tr}\{{\bf W}{\bf W}^H\}{\bf H}_{ir}^H{\bf H}_{ir} 
        +\sigma_{ar}^2\text{Tr}({\bf W}{\bf W}^H){\bf I}+\sigma_r^2{\bf I}.\label{useversion}
    \end{aligned}
\end{equation}
Based on (\ref{imcSIcp}) and (\ref{useversion}), the SR maximization problem under imperfect CSI is formulated as,
\begin{subequations}
\begin{align}
\min _{{\bf W}, {\bf \tilde \Theta},{\bf \tilde \Psi}} &\quad \frac{\left|{\bf I}_{N_e} +  {\bf \tilde H}_e({\bm {\tilde \Theta}}){\bf W}{\bf W}^H {\bf \tilde H}_e^H({\bm {\tilde \Theta}}){\bf J}_e^{-1}\right|}{\left|{\bf I}_{N_b} +  {\bf \tilde H}_b({\bm {\tilde \Theta}}){\bf W}{\bf W}^H {\bf \tilde H}_b^H({\bm {\tilde \Theta}}){\bf J}_b^{-1}\right|}, \\
 \text{s.t.} &\quad \text{(\ref{reofor41}), (\ref{reofor42}), (\ref{reofor43}), and (\ref{reofor44}).}
\end{align}
\label{problemImperCSI}%
\end{subequations}
Compared to the perfect CSI problem (\ref{reformualteobnew4}), problem (\ref{problemImperCSI}) differs only in the objective function by including the error covariance terms ${\bf J}_a$ and ${\bf J}_b$. 

\subsection{An Extension of P-PRCGD to Handle Imperfect CSI}
Since problem (\ref{problemImperCSI}) shares the same variables and constraints as its perfect CSI counterpart, the underlying PRM $\mathcal{M}$ and its tangent space $\mathcal{T}_{\bf \Upsilon} \mathcal{M}$ remain unchanged. Therefore, problem (\ref{problemImperCSI}) can also be reformulated using the AL method, similar to (\ref{nowopt111}),
\begin{equation}
  \min_{\{{\bf W}, {\bf\tilde \Theta},{\bf \tilde\Psi}\}\in\mathcal{M}} \quad  \left\{\begin{array}{l} f_{\text{Im}}({\bf W}, {\bf\tilde \Theta},{\bf \tilde\Psi})\\ + \frac{1}{2\rho}\sum_{g=1}^G\left\| {\bf \Psi}_g-{\bf \Theta}_g  \right\|^2\\+\sum_{g=1}^G\Re\left\{\text{Tr}\left\{{\bf \Phi}_g^H({\bf \Psi}_g-{\bf \Theta}_g)\right\}\right\}\end{array} \right\},\label{newimCSIpro}
\end{equation}
where $f_{\text{Im}}({\bf W}, {\bf\tilde \Theta}, {\bf \tilde\Psi})$ denotes the objective function of problem (\ref{problemImperCSI}). This reformulation reveals that, to adapt problem (\ref{problemImperCSI}) to the P-PRCGD method, only the Euclidean gradient of the objective function, calculated in step 4 of Algorithm \ref{alg:1}, requires modification. The updated Euclidean gradient of problem (\ref{problemImperCSI}) can be derived similarly to that in \textit{Lemma \ref{lemma41}} (details are omitted for brevity). By substituting this updated gradient into the update rule in (\ref{calculReg2}) within step 4 of Algorithm \ref{alg:1}, while keeping the other steps unchanged, the P-PRCGD method can be directly extended to solve problem (\ref{newimCSIpro}). Moreover, since the extension only modifies the objective function and its gradient, the previously established stationary convergence result directly applies to problem (\ref{newimCSIpro}).

\section{Numerical Results}
In this section, we present numerical results to demonstrate the enhanced PLS achieved with the assistance of BD-RIS in a MIMO system. Specifically, we examine the proposed method for fully-connected BD-RIS (denoted as FC-RIS) with $G=1$, as well as group-connected BD-RIS (denoted as GC-RIS). For comparison, the following schemes are considered:
\begin{itemize}
\item \textbf{UPPER-FC-RIS}: An upper bound obtained by solving (\ref{overallproblem}) with $R_b(\cdot)$ maximized and $R_e(\cdot)$ set to zero.  
\item \textbf{AO-FC-RIS/AO-GC-RIS}: Solving (\ref{reformualteobnew4}) using the AO algorithm. In each iteration, $\bf W$ is derived based on \cite{CHENG2024109571}, while $\bf \Theta$ is optimized using the CGD algorithm in Euclidean space, followed by a direct projection to maintain its constraints \cite{fang2023low}.
\item \textbf{D-RIS}: Enhancing PLS in a MIMO system with the assistance of a D-RIS \cite{hong2020artificial}.
\item \textbf{R-FC-RIS}: Enhancing PLS in a MIMO system with the assistance of a fully connected BD-RIS, optimizing only $\bf W$ \cite{CHENG2024109571}, while $\bf \Theta$ is randomly generated to maintain its constraints without optimization.
\item \textbf{WO-RIS}: Enhancing PLS in a MIMO system by optimizing only $\bf W$ without RIS assistance \cite{CHENG2024109571}.
\end{itemize}

\begin{figure}[t]
  \begin{center}
  \includegraphics[width=3in]{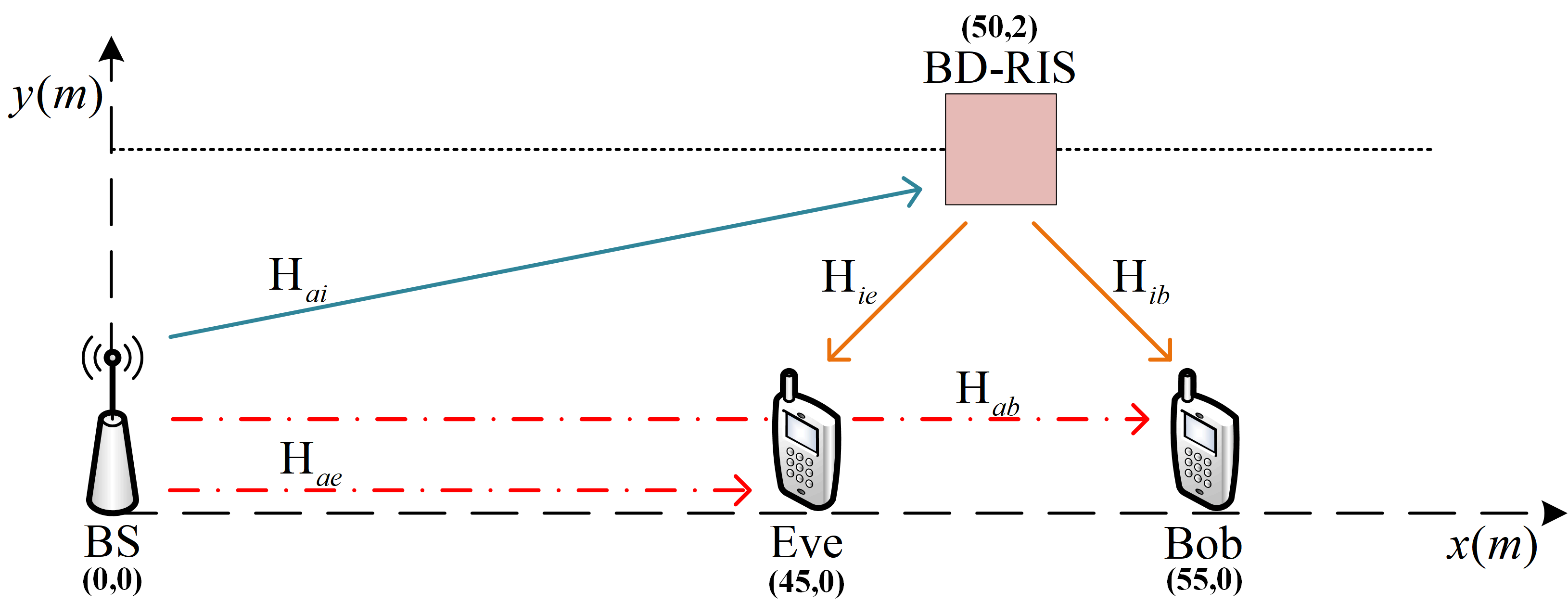}\\
  \caption{BD-RIS assisted MIMO communication scenario in simulation.}\label{MIMO_config}
  \end{center}
\end{figure}

Following \cite{asaad2022secure,hong2020artificial,dong2020enhancing}, we model the channels as comprising both large-scale and small-scale fading components, as shown in Fig. \ref{MIMO_config}. The large-scale path loss is given by $L(d) = C_0 (d/D_0)^{-\zeta}$, where \( d = \sqrt{(x_r - x_t)^2+(y_r - y_t)^2 } \) denotes the distance between the transmitter and receiver, with \( (x_t, y_t) \) and \( (x_r, y_r) \) representing their respective locations. Here, \( \zeta \) is the path loss exponent, and \( C_0 = -30 \) dB represents the path loss at the reference distance \( D_0 = 1 \) m.

For the direct channels ${\bf H}_{ae}$ and ${\bf H}_{ab}$, the small-scale fading is modeled as a complex Gaussian distribution. For the BD-RIS-related channels \( \mathbf{H}_{ai} \), \( \mathbf{H}_{ib} \), and \( \mathbf{H}_{ie} \), small-scale fading follows a Rician distribution, given by,    
\begin{equation}
    \mathbf{\tilde{H}} = \sqrt{\tfrac{\kappa}{1+\kappa}} \mathbf{\tilde{H}}^{\text{LOS}} + \sqrt{\tfrac{1}{1+\kappa}} \mathbf{\tilde{H}}^{\text{NLOS}},
\end{equation}  
where \( \mathbf{\tilde{H}}^{\text{LOS}} = \mathbf{a}_r(\psi_r) \mathbf{a}_t^H(\psi_t) \) represents the line-of-sight (LoS) component, where \( \mathbf{a}_r(\psi_r) \) and \( \mathbf{a}_t(\psi_t) \) are the steering vectors of the receive and transmit arrays, respectively, expressed as \( \mathbf{a}_j = [1, e^{j\pi\sin(\psi_j)}, \dots, e^{j\pi(M_j-1)\sin(\psi_j)}], \forall j \in \{r,t\} \), with the angles of departure (AoD) and arrival (AoA) given by \( \psi_t = \tan^{-1}(\frac{y_r - y_t}{x_r - x_t}) \) and \( \psi_r = \pi - \psi_t \), respectively. \(\mathbf{\tilde{H}}^{\text{NLOS}} \) is the non-line-of-sight (NLOS) component follows a complex Gaussian distribution. 

Unless otherwise specified, the parameters are set as follows: Alice, Bob, and Eve are equipped with \( N_t = 24 \), \( N_b = 4 \), and \( N_e = 2 \) antennas, respectively. The BD-RIS consists of \( M = 80 \) REs, and the number of data streams is \( N_s = 2 \). The total transmit power is \( P = 0 \) dBW, while the noise power at Bob and Eve is \( \sigma_b^2 = \sigma_e^2 = -40 \) dBm. The group-connected BD-RIS comprises \( G = 4 \) groups.  

The path loss exponents for the channels \( \mathbf{H}_{ai} \), \( \mathbf{H}_{ib} \), \( \mathbf{H}_{ie} \), \( \mathbf{H}_{ab} \), and \( \mathbf{H}_{ae} \) are set to \( \zeta_{ai} = 2.2 \), \( \zeta_{ib} = 2.5 \), \( \zeta_{ie} = 2.5 \), \( \zeta_{ab} = 3.5 \), and \( \zeta_{ae} = 3.5 \), respectively. The Rician factor is \( \kappa = 5 \). The coordinates of Alice, BD-RIS, Eve, and Bob are given as (0,0) m, (50,2) m, (45,0) m, and (55,0) m, respectively. The variables $\mathbf{W}$, $\mathbf{\Theta}$, and $\mathbf{\Psi}$ are initialized with random matrices, which are then projected onto their respective constraints. All results are averaged over 200 independent realizations and obtained on a PC with a 3.8 GHz Intel Core Ultra 7 155H processor.

\begin{figure}[t]
  \begin{center}
  \includegraphics[width=3in]{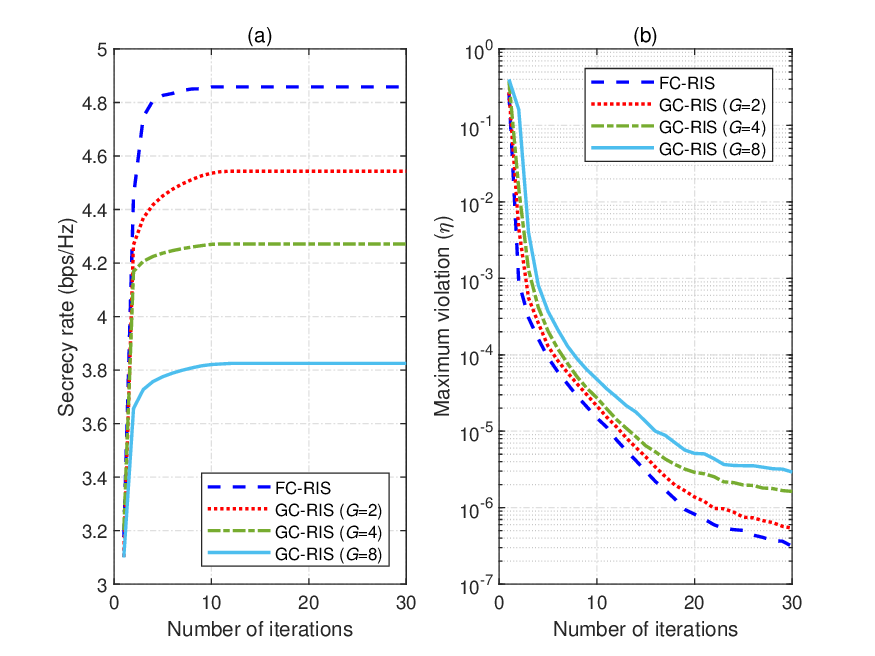}\\
  \caption{Convergence performance of the P-PRCGD method.}\label{Convergce}
  \end{center}
\end{figure}

\begin{figure}[t]
  \begin{center}
  \includegraphics[width=3in]{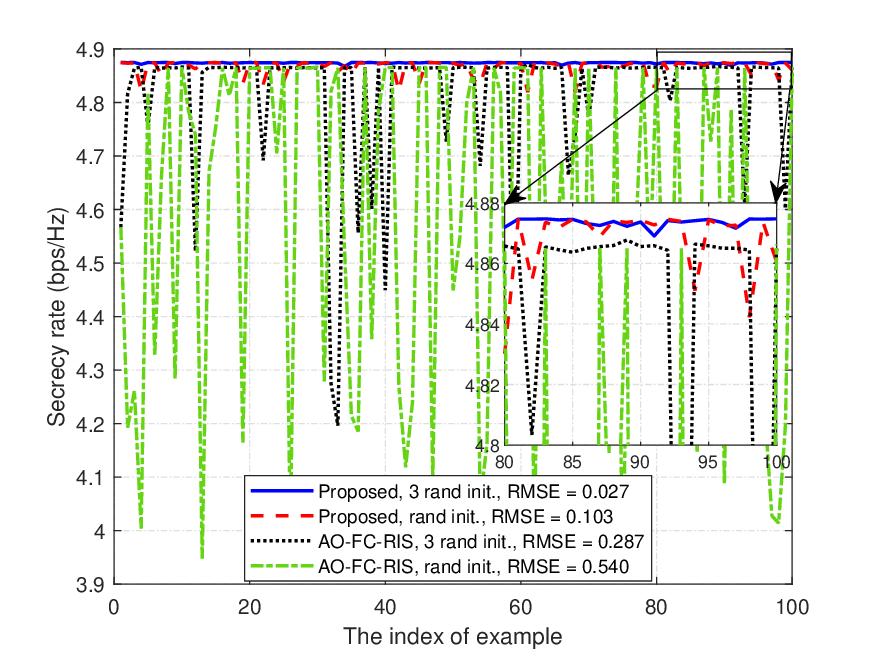}\\
  \caption{Comparison of secrecy rate between different architectures over 100 random initial points.}\label{Rum_ini}
  \end{center}
\end{figure}

Fig. \ref{Convergce} illustrates the convergence behavior of the proposed P-PRCGD method. As shown in Fig. \ref{Convergce}(a), the SR for all configurations increases monotonically, with FC-RIS outperforming GC-RIS due to its greater DoFs. The performance of GC-RIS predictably degrades as the number of groups ($G$) increases, which reduces the available DoFs. Fig. \ref{Convergce}(b) complements this by showing the monotonic decrease of the maximum constraint violation, $\eta$, confirming that the algorithm converges to a feasible solution. It is also observed that a larger $G$ results in a slower convergence rate for $\eta$, as optimizing a greater number of auxiliary variables makes constraint enforcement more challenging.

Fig. \ref{Rum_ini} evaluates the sensitivity of the proposed method to initialization, in comparison with the AO-FC-RIS algorithm, by assessing their SR performance over 100 randomly generated examples. Stability is quantified using the root mean square error (RMSE) of the achieved SRs, defined as, $\text{RMSE} = \sqrt{\frac{1}{K} \sum_{k=1}^{K} \left(SR_k - \bar{SR}\right)^2}$, where $SR_k$ is the SR is the SR in $k$-th trial, $\bar{SR}$ is the average SR, and $K=100$ is the total number of experiments. Lower RMSE values indicate higher robustness to initialization. The proposed method demonstrates superior stability, achieving an RMSE of 0.103, substantially lower than the 0.540 RMSE of the AO-FC-RIS method. This improved stability is attributed to the guaranteed convergence to a stationary point, which consistently guides the algorithm toward high-quality solutions. Furthermore, we evaluated a multi-start strategy in which the best result among three random initializations is selected. This approach further enhances stability, reducing the RMSE of the proposed method to 0.027, compared to 0.287 for the AO-FC-RIS method under the same strategy.

\begin{table}[t]
  \centering
  \caption{Computational Complexity Comparison of Different Methods.}
  \label{tab:complexity_comparison_separated}
  \begin{tabular}{ll}
    \toprule 
    \textbf{Methods} & \textbf{Computational Complexity} \\
    \midrule 
    AO-FC/GC-RIS \cite{CHENG2024109571,fang2023low} & $\mathcal{O}(IJ(M\tilde{M}^2 + N_t\tilde{M} + N_t^2))$ \\
    D-RIS \cite{hong2020artificial} & $\mathcal{O}(IJ(M^2 + N_t^2))$ \\
    R-FC/WO-RIS \cite{CHENG2024109571} & $\mathcal{O}(JN_t^2)$ \\
    \midrule 
    \textbf{Proposed}  & $\mathcal{O}(IJ\tilde{M}^2N_t)$ \\
    \bottomrule 
  \end{tabular}
\end{table}

\begin{figure}[t]
  \begin{center}
  \includegraphics[width=3in]{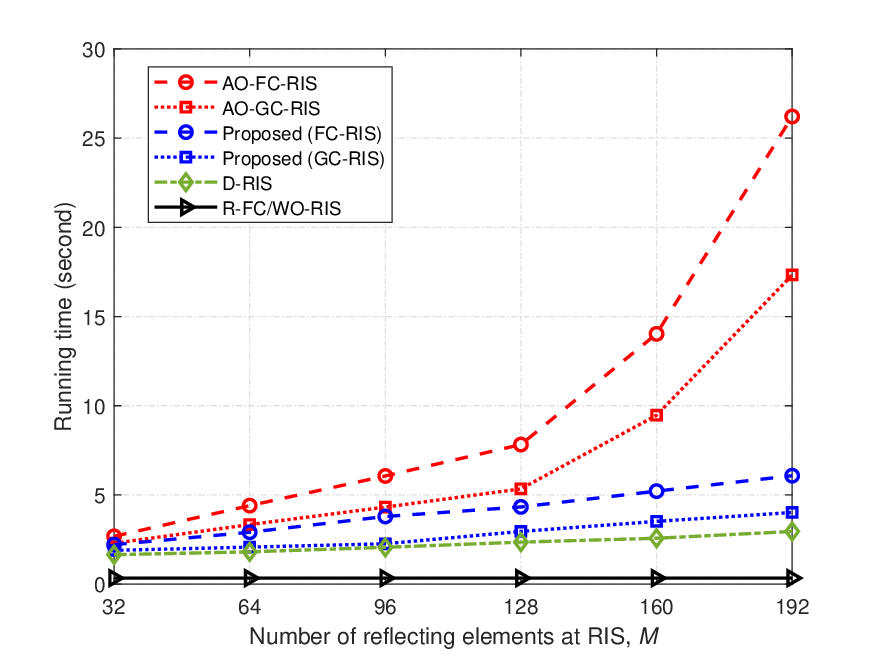}\\
  \caption{Comparison of running time between different architectures as the
number of REs at RIS ($M$) increases from 32 to 192.}\label{Rum_t}
  \end{center}
\end{figure}

To evaluate the computational efficiency of our proposed method, we first provide a theoretical complexity comparison in Table \ref{tab:complexity_comparison_separated}. The complexity of our method, \( \mathcal{O}(IJ\tilde{M}^2N_t) \), is notably lower than that of the AO-based benchmarks, as it avoids iterative sub-optimizations. While this is higher than the simpler D-RIS and WO-RIS schemes, it is a necessary trade-off for optimizing the additional DoFs offered by the BD-RIS architecture. This theoretical analysis is empirically corroborated by the running time results presented in Fig. \ref{Rum_t}. The results show a strong consistency with our analysis and, more importantly, highlight the superior scalability of our approach. As the number of REs ($M$) grows, the AO-based methods become computationally prohibitive, whereas our proposed P-PRCGD algorithm maintains a significantly more moderate and manageable running time, demonstrating its practical efficiency for large-scale systems.

\begin{figure}[t]
  \begin{center}
  \includegraphics[width=3in]{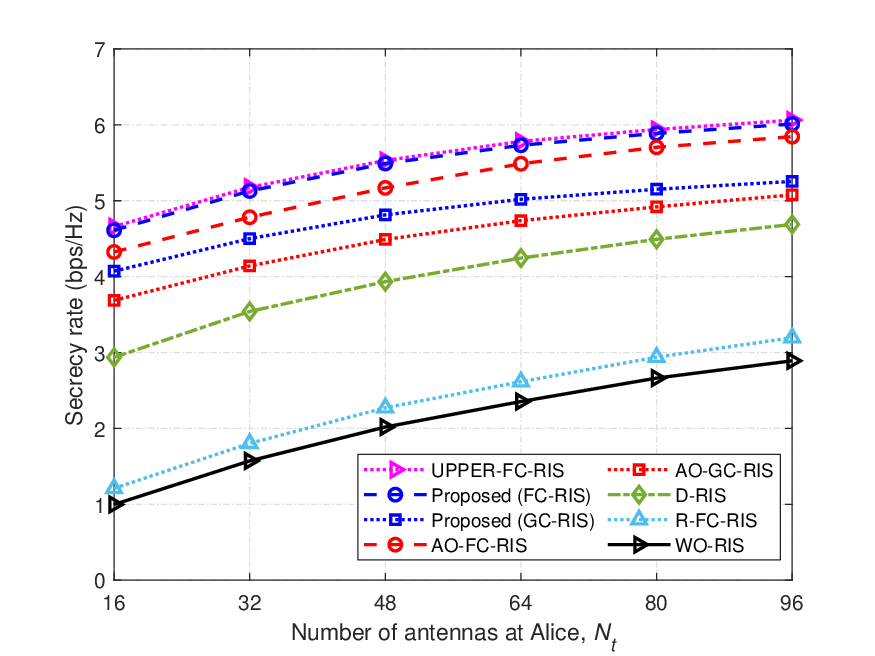}\\
  \caption{Comparison of secrecy rate between different architectures as the
number of antennas at the Alice ($N_t$) increases from 16 to 96.}\label{Dif_M}
  \end{center}
\end{figure}

\begin{figure}[t]
  \begin{center}
  \includegraphics[width=3in]{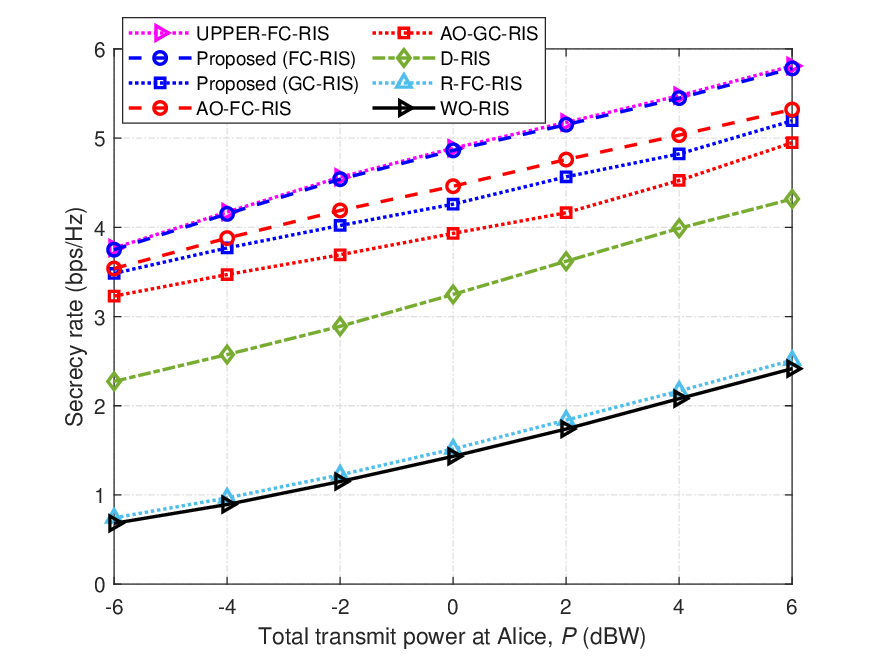}\\
  \caption{Comparison of secrecy rate between different architectures as the total transmit power at the Alice ($P$) increases from -6 dBW to 6 dBW.}\label{Dif_P}
  \end{center}
\end{figure}

Figs. \ref{Dif_M} and \ref{Dif_P} illustrate the SR performance as the number of transmit antennas ($N_t$) and the transmit power ($P$) vary. All schemes benefit from increases in $N_t$ and $P$ due to the expanded spatial DoFs and a larger power budget. Our proposed method consistently achieves the second-highest SR, performing very close to the theoretical upper bound. Specifically, at $N_t = 32$ or $P = 0$ dBW, our P-PRCGD algorithm with FC-RIS achieves an SR of 4.86 bps/Hz, representing a notable 34.7\% improvement over the conventional D-RIS. This performance is also 8.3\% higher than that of the AO-based method, highlighting the advantage of our Riemannian optimization approach in handling non-convex constraints without relaxation.

\begin{figure}[t]
  \begin{center}
  \includegraphics[width=3in]{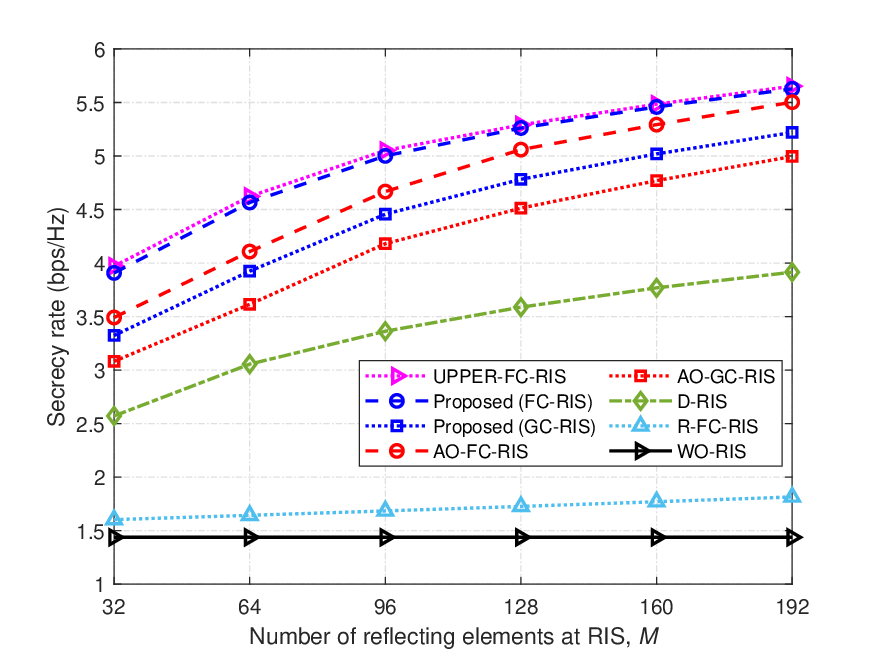}\\
  \caption{Comparison of secrecy rate between different architectures as the
number of REs at RIS ($M$) increases from 32 to 192.}\label{Dif_IRS}
  \end{center}
\end{figure}

\begin{figure}[t]
  \begin{center}
  \includegraphics[width=3in]{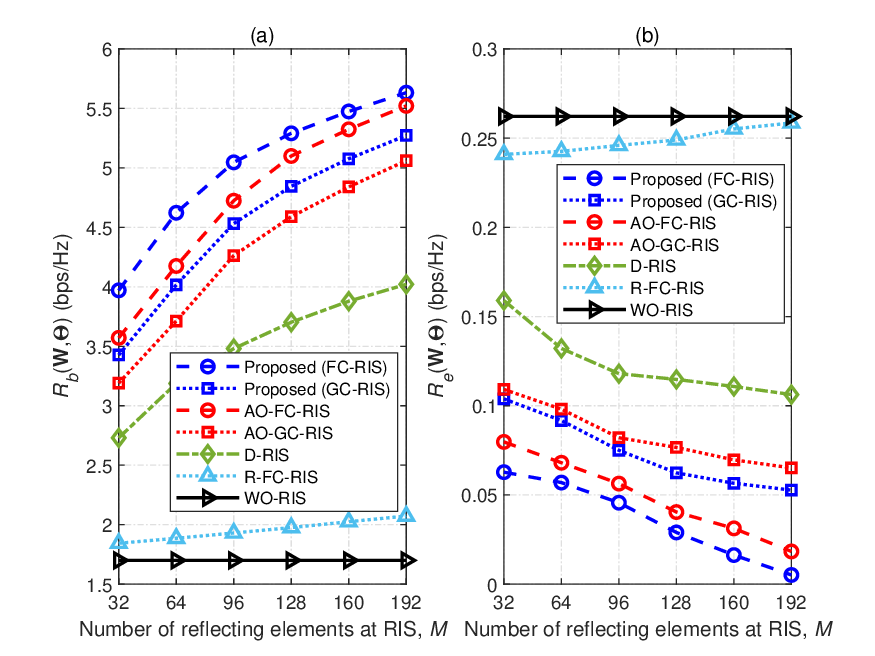}\\
  \caption{Comparison of channel capacities for (a) Bob ($R_b({\bf W},{\bm\Theta})$) and (b) Eve ($R_e({\bf W},{\bm\Theta})$) between different architectures as the number of REs at RIS ($M$) increases from 32 to 192.}\label{RbRe}
  \end{center}
\end{figure}

Fig. \ref{Dif_IRS} presents the SR performance versus the number of RIS elements, $M$. As $M$ increases, all RIS-assisted schemes exhibit improved SR, confirming that additional DoFs enhance system performance. Notably, our proposed method achieves the second-best performance, closely approaching the upper bound. Specifically, at $M = 64$, our P-PRCGD with FC-RIS attains an SR of 4.54 bps/Hz, outperforming AO-FC-RIS by 9.5\%, D-RIS by 34.1\%, and the unoptimized R-FC-RIS by 66.1\%. Similarly, our GC-RIS solution exceeds its AO counterpart by 7.9\%. These results validate that an optimized BD-RIS offers substantial SR improvements over both conventional and randomly configured RIS schemes.

To further analyze this SR improvement, Fig. \ref{RbRe} decomposes the performance into the individual channel capacities for Bob ($R_b$) and Eve ($R_e$). Our proposed BD-RIS method uniquely achieves a dual objective: it significantly enhances Bob's capacity while actively suppressing Eve's, a clear advantage over D-RIS and no-RIS scenarios. An important counter-example is the R-FC-RIS benchmark, where Eve's capacity undesirably increases with $M$. This phenomenon demonstrates that without sophisticated optimization, the extra DoFs from a BD-RIS can be inadvertently exploited by the eavesdropper. This highlights the necessity of our proposed algorithm to intelligently manage the BD-RIS and reliably translate its advanced capabilities into guaranteed SR gains.

\begin{figure}[t]
  \begin{center}
  \includegraphics[width=3in]{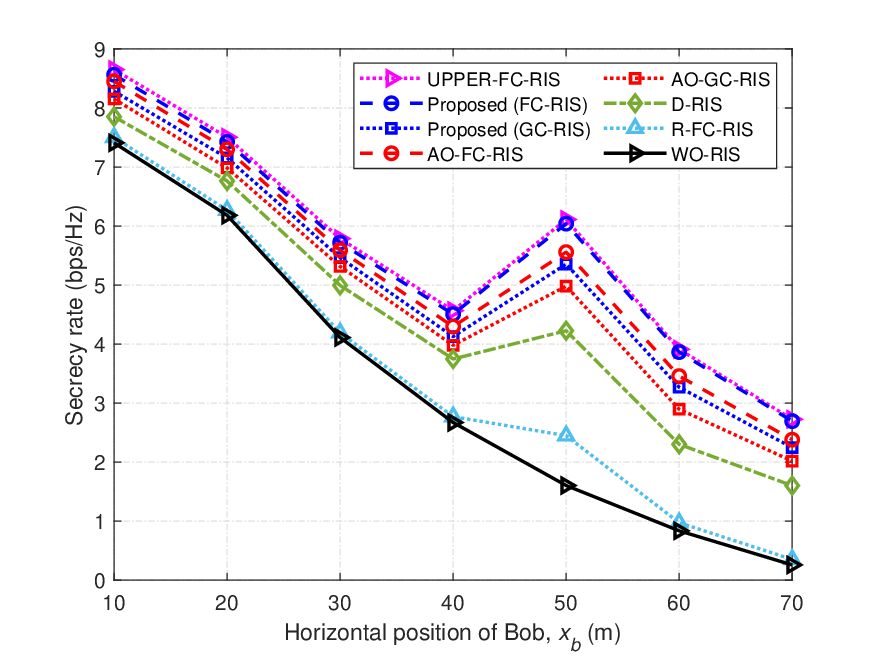}\\
  \caption{Comparison of secrecy rate between different architectures as the horizontal position of Bob ($x_b$) increases from 10 m to 70 m.}\label{Dif_Dis}
  \end{center}
\end{figure}

Fig. \ref{Dif_Dis} illustrates the SR performance as Bob's horizontal position, $x_b$, varies, while Eve and the RIS remain at fixed locations. As expected, the SR for all schemes generally decreases as the Alice–Bob distance increases due to path loss in the direct channel, $\mathbf{H}_{ab}$. However, a notable trend reversal occurs for RIS-assisted schemes in the range $x_b \in [40, 50]$, where the RIS-to-Bob channel, $\mathbf{H}_{ib}$, becomes dominant. This effect peaks at $x_b = 50$ m, where Bob is orthogonal to the RIS, creating optimal channel conditions. In this critical region, our proposed method demonstrates superior capability in manipulating the propagation environment. Specifically, the SR gain from $x_b = 40$ to $x_b = 50$ for our method is 34.2\% (FC-RIS) and 29.9\% (GC-RIS), significantly outperforming the gains achieved by the AO method (28.3\% and 25.1\%) and the D-RIS scheme (12.8\%). This highlights both the high efficiency of our algorithm and the advanced wave-shaping capabilities of the BD-RIS architecture.

\begin{figure}[t]
  \begin{center}
  \includegraphics[width=3in]{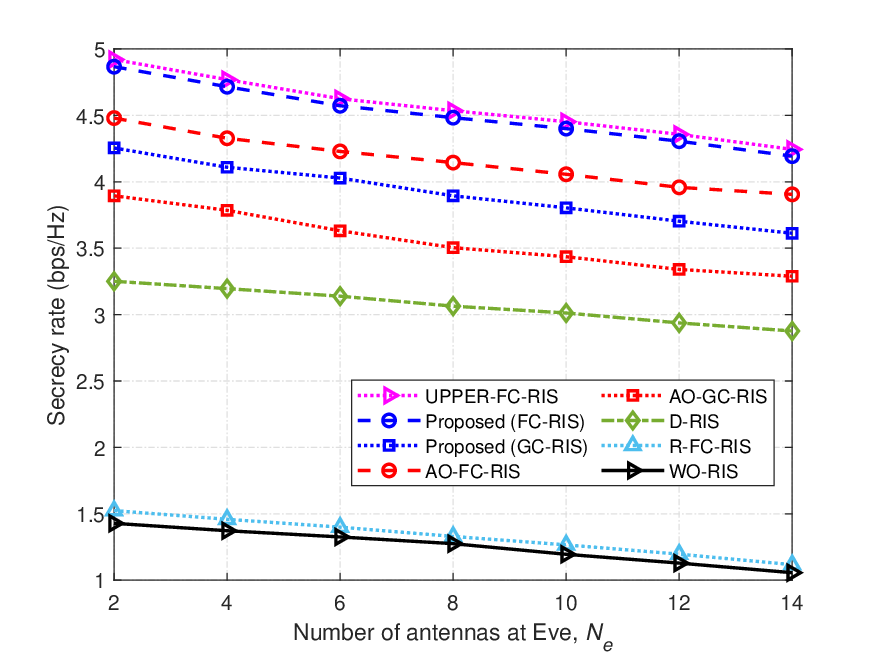}\\
  \caption{Comparison of secrecy rate between different architectures as the number of antennas at Eve ($N_e$) increases from 2 to 14.}\label{Dif_EVE}
  \end{center}
\end{figure}

\begin{figure}[t]
  \begin{center}
  \includegraphics[width=3in]{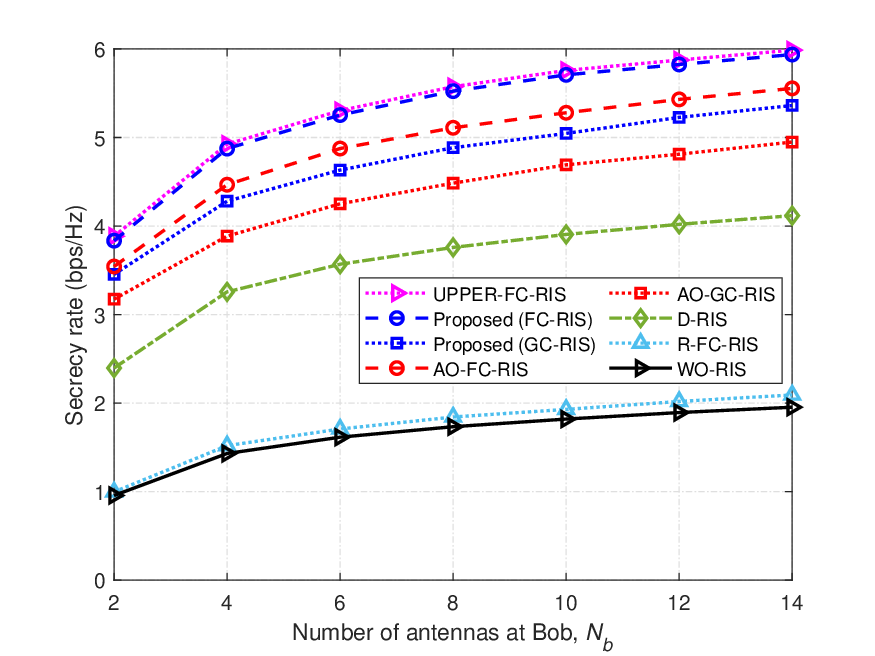}\\
  \caption{Comparison of secrecy rate between different architectures as the number of antennas at Bob ($N_b$) increases from 2 to 14.}\label{Dif_BOB}
  \end{center}
\end{figure}

Figs. \ref{Dif_EVE} and \ref{Dif_BOB} illustrate the impact of the number of receive antennas at Eve ($N_e$) and Bob ($N_b$) on the SR. As expected, the SR improves with increasing $N_b$ and degrades with increasing $N_e$. In both scenarios, our proposed P-PRCGD algorithm with BD-RIS consistently achieves the second-best performance, closely approaching the upper bound, while outperforming all benchmark schemes. For instance, compared to the conventional D-RIS, our method with FC-RIS and GC-RIS yields SR improvements of approximately 32\% and 22–23\%, respectively. More importantly, our approach demonstrates a consistent 7–10\% SR gain over the AO-based method, highlighting the superior efficiency of our Riemannian optimization framework.

\begin{figure}[t]
  \begin{center}
  \includegraphics[width=3in]{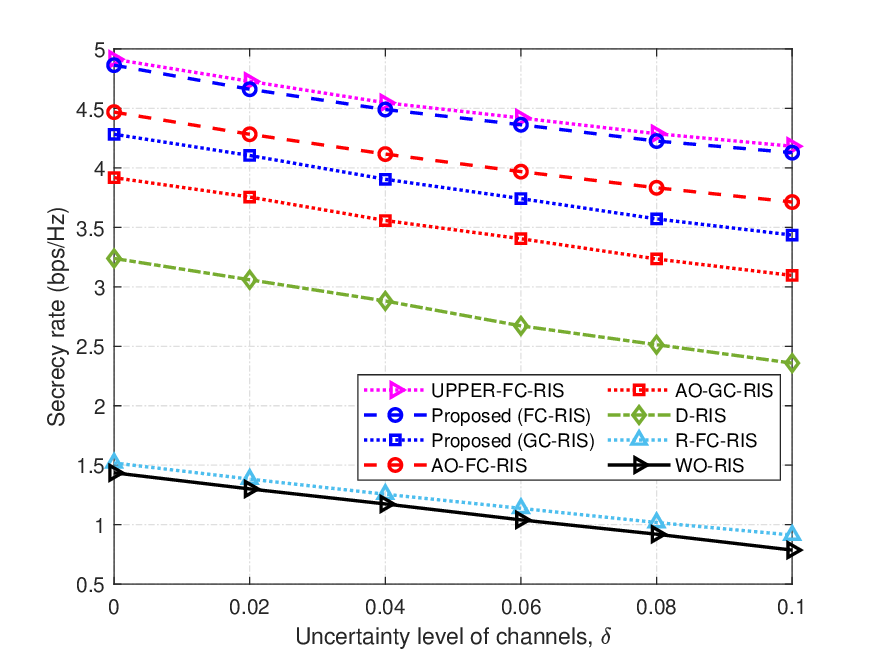}\\
  \caption{Comparison of secrecy rate between different architectures as the
uncertainty level of channels ($\delta$) increases from 0 to 0.1.}\label{Dif_Incsi}
  \end{center}
\end{figure}

Fig. \ref{Dif_Incsi} evaluates the robustness of all benchmark schemes to imperfect CSI. This robustness is quantified by the uncertainty level $\delta$, defined as: $\delta = \mathbb{E}\{\|\Delta\mathbf{H}\|^2\} / \sigma^2$, where $\Delta\mathbf{H}$ represents the CEEs for any of the channels as defined in (\ref{usenow1now}) and $\sigma^2$ is the corresponding variance from (\ref{simpleCSGs}). As expected, the SR of all methods degrades with increasing $\delta$, yet our proposed method demonstrates significantly greater resilience. As $\delta$ increases from 0 to 0.1, the SR degradation for our method with FC-RIS and GC-RIS is only 15.13\% and 19.80\%, respectively. This is substantially lower than the degradation observed for D-RIS (27.18\%), R-FC-RIS (39.95\%), and WO-RIS (45.26\%), highlighting the BD-RIS architecture's superior capability to mitigate the adverse effects of channel uncertainty. Furthermore, our P-PRCGD algorithm again outperforms the AO-based approach, whose SR degradation with FC-RIS (19.80\%) and GC-RIS (27.18\%) is noticeably higher. This result empirically confirms the superior robustness of our proposed framework in practical, non-ideal scenarios.

\section{Conclusion}
In this paper, we investigated the PLS of a BD-RIS-assisted MIMO communication system. We considered a scenario where confidential information was transmitted to legitimate users with the assistance of a BD-RIS to enhance security, while eavesdroppers attempted to intercept the signal. To maximize SR, we jointly optimized the beamforming at the transmitter and the REs at the BD-RIS, subject to total power and structural constraints. To address the non-convex SR maximization problem, we first reformulated it into a more tractable form by introducing an auxiliary variable to decouple the BD-RIS constraints. We then proposed a low-complexity P-PRCGD method, which integrated the AL method with the PMGD algorithm to solve the reformulated problem. Simulation results demonstrated the superior performance of the proposed method.

\begin{appendices} 
\section{proof of lemma \ref{lemma41}}
\label{appendixA}
To derive the Euclidean gradients \( \nabla_{\mathbf{W}} \mathcal{L}(\mathbf{W}) \), \( \nabla_{\mathbf{\Theta}_g} \mathcal{L}(\mathbf{\Theta}_g) \) for all \( g \), and \( \nabla_{\mathbf{\Psi}_g} \mathcal{L}(\mathbf{\Psi}_g) \) for all \( g \), we first reformulate the objective function in (\ref{new1111}) as follows, omitting the constant term for clarity,
\begin{equation}
    \mathcal{L}({\bf W}, {\bf\tilde \Theta},{\bf \tilde\Psi})=\frac{f_e}{f_b}+\frac{1}{2\rho}\sum_{g=1}^G\|{\bf \Psi}_g-{\bf \Theta}_g+\rho{\bf \Phi}_g \|^2. \label{useproofEg}
\end{equation}
where $f_e=|{\bf I}_{N_e} +  {\bf \tilde H}_e({\bm {\tilde \Theta}}){\bf W}{\bf W}^H {\bf \tilde H}_e^H({\bm {\tilde \Theta}})|$, $f_b=|{\bf I}_{N_b} +  {\bf \tilde H}_b({\bm {\tilde \Theta}}){\bf W}{\bf W}^H {\bf \tilde H}_b^H({\bm {\tilde \Theta}})|$. We now begin the proof.

\textit{Calculation of} \( \nabla_{{\bf W}} \mathcal{L}({\bf W}) \): By regarding other variables except for $\bf W$ as constant, and ignoring the constant term, the following is derived according to the quotient rule,
\begin{equation}
    \nabla_{{\bf W}} \mathcal{L}({\bf W}) = \frac{\nabla_{{\bf W}}f_e\cdot f_b-\nabla_{{\bf W}}f_b \cdot f_e}{(f_b)^2},\label{EDW}
\end{equation}
where,
\begin{equation}
        \nabla_{{\bf W}}f_r= 2  {\bf \tilde H}_r^H({\bf \Theta}_g) \text{adj}({\bf \hat F}_r){\bf \tilde H}_r({\bf \Theta}_g){\bf W},\quad r\in\{e,b\},
\end{equation}
where ${\bf \hat F}_r={\bf I}_{N_i} +  {\bf \tilde H}_r({\bm {\tilde \Theta}}){\bf W}{\bf W}^H {\bf \tilde H}_r^H({\bm {\tilde \Theta}})$, $\text{adj}({\bf \hat F}_i)$ is the adjugate matrix of ${\bf \hat F}_i$, and \( r \) indexes \( e \) (Eve) or \( b \) (Bob).

\textit{Calculation of} \( \nabla_{{\bf \Theta}_g} \mathcal{L}({\bf \Theta}_g), \forall g \): Similar as the derivation above, by regarding other variables as constant, the following can be obtained,
\begin{equation}
    \nabla_{{\bf \Theta}_g} \mathcal{L}({\bf \Theta}_g) = \frac{\nabla_{{{\bf \Theta}_g}}f_e\cdot f_b-\nabla_{{{\bf \Theta}_g}}f_b \cdot f_e}{(f_b)^2}-\frac{1}{\rho}({\bf \Psi}_g-{\bf \Theta}_g+\rho{\bf \Phi}_g)\label{EDTHETA}
\end{equation}
where,
\begin{equation}
   \nabla_{{\bf \Theta}_g}f_r= 2\sqrt{\frac{P}{\sigma_e^2}}({\bf \tilde H}_{ir}^g)^H\text{adj}({\bf \hat F}_r){\bf \tilde H}_r({\bf \Theta}_g){\bf W}{\bf W}^H({\bf \tilde H}_{ai}^g)^H,    
\end{equation}
where $\forall g \in G$, and $r$ is defined similar as above.

\textit{Calculation of} \(  \nabla_{\mathbf{\Psi}_g} \mathcal{L}(\mathbf{\Psi}_g), \forall g \): The objective function, excluding the constant term for \( \mathbf{\Psi}_g \), is a quadratic term, and its Euclidean gradient is straightforward to compute:
\begin{equation}
\nabla_{\mathbf{\Psi}_g} \mathcal{L}(\mathbf{\Psi}_g) = \frac{1}{\rho} (\mathbf{\Psi}_g - \mathbf{\Theta}_g + \rho \mathbf{\Phi}_g), \quad \forall g.\label{EDPSI}
\end{equation}
This completes the proof. $\hfill\blacksquare$

\section{Proof of Theorem \ref{theorem1}}
\label{appendixB}
To begin with, we show that the Riemannian Zoutendijk condition hold for \(\{{\bf \Upsilon}^q\}\), based on the \textit{Lemma \ref{lemmaAPB1}} as follows. 

\begin{lemma} \label{lemmaAPB1}
Suppose that a sequence $\{{\bf \Upsilon}^q\}$ on a Riemannian manifold $\mathcal{M}$ is generated with a retraction in (\ref{useforretrac}). Suppose also that each search direction ${\bf D}^q$ satisfies $\langle\text{grad}_{\mathcal{M}} \mathcal{L}({\bf \Upsilon}^{q}),{\bf D}^{q}  \rangle<0$ and each step size $\alpha^q$ satisfies the Wolfe conditions in (\ref{RWCONdi}). If (\ref{Lipschitzc}) is satisfied, then the following series converges,
\begin{equation}
    \sum_{q=1}^\infty\frac{\langle\text{grad}_{\mathcal{M}} \mathcal{L}({\bf \Upsilon}^q), {\bf D}^{q} \rangle^2}{\| {\bf D}^{q} \|^2}<\infty,\label{now1123}
\end{equation}
\end{lemma}
\textit{Proof}: See \textit{Theorem 2} in \cite{ring2012optimization}. $\hfill\blacksquare$

Next, we demonstrate that the following key inequality holds for \( \{ {\bf \Upsilon}^q \} \),
\begin{equation}
    -\frac{1}{1-\sigma_2}\le\frac{\langle\text{grad}_{\mathcal{M}} \mathcal{L}({\bf \Upsilon}^q), {\bf D}^{q} \rangle}{\| \text{grad}_{\mathcal{M}} \mathcal{L}({\bf \Upsilon}^q) \|^2}\le\frac{2\sigma_2-1}{1-\sigma_2}.\label{keyineq2}
\end{equation}
We proof it by the mathematical induction. For $q=1$, \( {\bf D}^1 = -\text{grad}_{\mathcal{M}} \mathcal{L}({\bf \Upsilon}^1) \), and it is evident that inequality (\ref{keyineq2}) holds. Now, suppose that inequality (\ref{keyineq2}) holds for $q\ge1$. We have,
\begin{subequations}
    \begin{align}
        &\frac{\langle\text{grad}_{\mathcal{M}} \mathcal{L}({\bf \Upsilon}^{q+1}), {\bf D}^{q+1} \rangle}{\| \text{grad}_{\mathcal{M}} \mathcal{L}({\bf \Upsilon}^{q+1}) \|^2}\notag\\&=\frac{\left\langle\begin{array}{l}\quad\quad\quad\quad\quad\quad\text{grad}_{\mathcal{M}} \mathcal{L}({\bf \Upsilon}^{q+1}),\\-\text{grad}_{\mathcal{M}} \mathcal{L}({\bf \Upsilon}^{q+1})+\beta^{q+1} \text{Tran}_{{{\bf \Upsilon}^{q}} \to {{\bf \Upsilon}^{q+1}}}({\bf D}^{q})  \end{array}\right\rangle}{\| \text{grad}_{\mathcal{M}} \mathcal{L}({\bf \Upsilon}^q) \|^2}\label{nownow1}\\&=-1+\frac{\langle\text{grad}_{\mathcal{M}} \mathcal{L}({\bf \Upsilon}^{q+1}), \text{Tran}_{{{\bf \Upsilon}^{q}} \to {{\bf \Upsilon}^{q+1}}}({\bf D}^{q}) \rangle}{\| \text{grad}_{\mathcal{M}} \mathcal{L}({\bf \Upsilon}^{q+1}) \|^2},\label{nownow2}
    \end{align}
\end{subequations}
where the first equality (\ref{nownow1}) follows from (\ref{desRe}), and the second equality (\ref{nownow2}) follows from (\ref{conjgateparemae}). Then, combining (\ref{refstepsize}) and (\ref{nownow2}), we obtain,
\begin{equation}
\begin{aligned}
    -1+\sigma_2\frac{\langle\text{grad}_{\mathcal{M}} \mathcal{L}({\bf \Upsilon}^q), {\bf D}^{q} \rangle}{\| \text{grad}_{\mathcal{M}} \mathcal{L}({\bf \Upsilon}^q) \|^2}&\le\frac{\langle\text{grad}_{\mathcal{M}} \mathcal{L}({\bf \Upsilon}^{q+1}), {\bf D}^{q+1} \rangle}{\| \text{grad}_{\mathcal{M}} \mathcal{L}({\bf \Upsilon}^{q+1}) \|^2}\\&\le-1-\sigma_2\frac{\langle\text{grad}_{\mathcal{M}} \mathcal{L}({\bf \Upsilon}^q), {\bf D}^{q} \rangle}{\| \text{grad}_{\mathcal{M}} \mathcal{L}({\bf \Upsilon}^q) \|^2}.\label{usenowproof3}
\end{aligned}    
\end{equation}
Applying the induction hypothesis (\ref{keyineq2}) for $q$, we conclude that inequality (\ref{keyineq2}) holds for $q+1$, thereby completing the proof.

Based on (\ref{now1123}) and (\ref{keyineq2}), we are now ready to complete the proof. To begin with, from (\ref{desRe}), we have,
\begin{subequations}
\begin{align}
   & \|{\bf D}^q\|^2=\| -\text{grad}_{\mathcal{M}} \mathcal{L}({\bf \Upsilon}^q)+\beta^q \text{Tran}_{{{\bf \Upsilon}^{q-1}} \to {{\bf \Upsilon}^q}}({\bf D}^{q-1})\|^2\label{nowproofcon1}\\&=\|\text{grad}_{\mathcal{M}} \mathcal{L}({\bf \Upsilon}^q)\|^2+(\beta^q)^2\|\text{Tran}_{{{\bf \Upsilon}^{q-1}} \to {{\bf \Upsilon}^q}}({\bf D}^{q-1})\|^2\notag\\&\quad\quad+2\beta^q|\langle \text{grad}_{\mathcal{M}} \mathcal{L}({\bf \Upsilon}^q), \text{Tran}_{{{\bf \Upsilon}^{q-1}} \to {{\bf \Upsilon}^q}}({\bf D}^{q-1})\rangle|\\&\le\|\text{grad}_{\mathcal{M}} \mathcal{L}({\bf \Upsilon}^q)\|^2+(\beta^q)^2\|{\bf D}^{q-1}\|^2\notag\\&\quad\quad+\tfrac{2\sigma_2}{1-\sigma_2}\beta^q\| \text{grad}_{\mathcal{M}} \mathcal{L}({\bf \Upsilon}^{q-1})\|^2\label{nowproofcon2}\\&=\|\text{grad}_{\mathcal{M}} \mathcal{L}({\bf \Upsilon}^q)\|^2+\tfrac{|| \text{grad}_{\mathcal{M}} \mathcal{L}({\bf \Upsilon}^q)||^4}{|| \text{grad}_{\mathcal{M}} \mathcal{L}({\bf \Upsilon}^{q-1})||^4}\|{\bf D}^{q-1}\|^2\notag\\&\quad\quad+\tfrac{2\sigma_2}{1-\sigma_2}\| \text{grad}_{\mathcal{M}} \mathcal{L}({\bf \Upsilon}^{q})\|^2\label{nowproofcon3}\\&=\tfrac{1+\sigma_2}{1-\sigma_2}\| \text{grad}_{\mathcal{M}} \mathcal{L}({\bf \Upsilon}^{q})\|^2+\tfrac{|| \text{grad}_{\mathcal{M}} \mathcal{L}({\bf \Upsilon}^q)||^4}{|| \text{grad}_{\mathcal{M}} \mathcal{L}({\bf \Upsilon}^{q-1})||^4}\|{\bf D}^{q-1}\|^2
\end{align} 
\label{proofnownewnew1}%
\end{subequations}
where the equality in (\ref{nowproofcon1}) follows from (\ref{desRe}), and the inequality in (\ref{nowproofcon2}) follows from the fact that \( \| \text{Tran}_{{{\bf \Upsilon}^{q-1}} \to {\bf \Upsilon}^q}({\bf D}^{q-1}) \| \leq \| {\bf D}^{q-1} \| \). Additionally, we have,
\begin{equation}
\begin{aligned}
    |\langle \text{grad}_{\mathcal{M}} \mathcal{L}({\bf \Upsilon}^q), &\text{Tran}_{{{\bf \Upsilon}^{q-1}} \to {{\bf \Upsilon}^q}}({\bf D}^{q-1})\rangle|\\&\le-\sigma_2\langle\text{grad}_{\mathcal{M}} \mathcal{L}({\bf \Upsilon}^{q}),{\bf D}^{q}  \rangle\\&\le \frac{\sigma_2}{1-\sigma_2}\| \text{grad}_{\mathcal{M}} \mathcal{L}({\bf \Upsilon}^{q-1})\|^2,
\end{aligned}    
\end{equation}
which follows from equations (\ref{refstepsize}) and (\ref{keyineq2}). The equality in (\ref{nowproofcon3}) follows from (\ref{conjgateparemae}). The successive application of the inequality (\ref{proofnownewnew1}) together with (\ref{conjgateparemae}) yields the following result,
\begin{equation}
    \begin{aligned}
        &\|{\bf D}^q\|^2\\=&\sigma\left(\begin{array}{l}\| \text{grad}_{\mathcal{M}} \mathcal{L}({\bf \Upsilon}^{q})\|^2+(\beta^q)^2\| \text{grad}_{\mathcal{M}} \mathcal{L}({\bf \Upsilon}^{q-1})\|^2 \\+ \cdots+(\beta^q)^2(\beta^{q-1})^2 \cdots (\beta^2)^2\| \text{grad}_{\mathcal{M}} \mathcal{L}({\bf \Upsilon}^{1})\|^2\end{array}\right)\\&+(\beta^q)^2(\beta^{q-1})^2 \cdots (\beta^1)^2 \|{\bf D}^0\|^2\\<&\sigma\| \text{grad}_{\mathcal{M}} \mathcal{L}({\bf \Upsilon}^{q})\|^4\sum_{j=0}^q\|\text{grad}_{\mathcal{M}} \mathcal{L}({\bf \Upsilon}^{j})\|^{-2}
    \end{aligned}
\end{equation}
where $\sigma=\frac{1+\sigma_2}{1-\sigma_2}$. We prove the convergence by contradiction. If we assume $\lim_{q \to \infty} \inf \|\text{grad}_{\mathcal{M}} \mathcal{L}({\bf \Upsilon}^q)\| \ne  0$, meaning that there exist a $\vartheta >0$, such that $\|\text{grad}_{\mathcal{M}} \mathcal{L}({\bf \Upsilon}^q)\| \ge  0$ for all $q$. Then, we have,
\begin{equation}
    \|{\bf D}^q\|^2\le\sigma\|\text{grad}_{\mathcal{M}} \mathcal{L}({\bf \Upsilon}^{q})\|^{4}\frac{q}{\vartheta^2}.
\end{equation}
This implies that,
\begin{equation}
    \sum_{q=1}^{\infty}\frac{\|\text{grad}_{\mathcal{M}} \mathcal{L}({\bf \Upsilon}^{q})\|^{4}}{\|{\bf D}^q\|^2}\ge\sigma\vartheta^2\sum_{q=1}^{\infty}\frac{1}{q}=\infty.\label{proofinfty}
\end{equation}
However, the inequality (\ref{now1123}) together with (\ref{keyineq2}) implies that $\sum_{q=1}^{\infty}\frac{\|\text{grad}_{\mathcal{M}} \mathcal{L}({\bf \Upsilon}^{q})\|^{4}}{\|{\bf D}^q\|^2}\le\infty$, which contradicts (\ref{proofinfty}). Therefore, we obtain the result,
\begin{equation}
    \lim_{q \to \infty} \inf \|\text{grad}_{\mathcal{M}} \mathcal{L}({\bf \Upsilon}^q)\| = 0,\label{finalproofth1}
\end{equation}
which ensures that Algorithm \ref{alg:1} converges to a stationary point of problem (\ref{new1111}). This completes the proof.$\hfill\blacksquare$

\section{Proof of Theorem \ref{theorem2}}
\label{appendixC}
For clarity, the augmented Lagrangian function in (\ref{LRfuncu}) is equivalent to,
\begin{equation}
    \mathcal{L}({\bf \Upsilon}) = f({\bf \Upsilon}) + \frac{1}{2\rho} \sum_{g=1}^G \| h_g({\bf \Upsilon}) + \rho {\bf \Phi}_g \|^2,
\end{equation}
where \( {\bf \Upsilon} \) is defined in (\ref{variablecomb}), and \( h_g({\bf \Upsilon}) = {\bf \Psi}_g - {\bf \Theta}_g \), since \( {\bf \Theta}_g \) is also an element of \( {\bf \Upsilon} \). We now begin the proof.

Without loss of generality, we assume that \( {\bf \Upsilon}^p \) converges to \( {\bf \Upsilon}^* \) (if not, we can restrict to a convergent subsequence of \( \{{\bf \Upsilon}^p\} \)). Since the LICQ condition \cite{boyd2004convex} is satisfied at \( {\bf \Upsilon}^* \), and by the continuity of the gradients of \( \{h_g({\bf \Upsilon})\}_{g=1}^G \), the tangent vectors (i.e., the Riemannian gradients) \( \{\text{grad}_\mathcal{M} h_g({\bf \Upsilon}^p)\}_{g=1}^G \) are linearly independent for all \( p \), i.e.,
\begin{equation}
    \left\|\sum_{g=1}^G \text{grad}_\mathcal{M} h_g({\bf \Upsilon}^*)\right\| \neq 0.\label{inequavutLIC}
\end{equation}
Define \( \mathcal{E}^p = \max\{\| {\bf \Phi}_g^{p} \|_{\infty}, \forall g\} \). We will discuss separately the cases when \( \mathcal{E}^p \) is bounded and when it is unbounded. If \( \mathcal{E}^p \) is bounded, denote the limit points of \( {\bf \Phi}_g^{p} \) as \( {\bf \Phi}_g^{*} \). Let,
\begin{equation}
    {\bf T} = \text{grad}_\mathcal{M} f({\bf \Upsilon}^*) + \sum_{g=1}^G {\bf \Phi}_g^{*} \, \text{grad}_\mathcal{M} h_g({\bf \Upsilon}^*).\label{refor2su}
\end{equation}
To prove that \( {\bf T} = 0 \), we compare it with a similar vector defined at \( {\bf \Upsilon}^p \) for large \( p \), and take the limit as \( p \to \infty \). Since directly comparing tangent vectors in the tangent spaces \( {\mathcal T}_{{\bf \Upsilon}^p} \mathcal{M} \) and \( {\mathcal T}_{{\bf \Upsilon}^*} \mathcal{M} \) at points \( {\bf \Upsilon}^p \) and \( {\bf \Upsilon}^* \) is challenging, we use parallel transport to map all tangent vectors to \( {\mathcal T}_{{\bf \Upsilon}^*} \mathcal{M} \) at \( {\bf \Upsilon}^* \). By applying the triangle inequality, (\ref{refor2su}) can be reformulated as,
\begin{equation}
\begin{aligned}
    \|{\bf T}\| \le& \left\|\begin{array}{l} \text{grad}_\mathcal{M} f({\bf \Upsilon}^*)-\text{Tran}_{{\bf \Upsilon}^p \to {\bf \Upsilon}^*}\text{grad}_\mathcal{M} f({\bf \Upsilon}^p)\\+\sum_{g=1}^G {\bf \Phi}_g^{*} ( \text{grad}_\mathcal{M} h_g({\bf \Upsilon}^*) \\-\text{Tran}_{{\bf \Upsilon}^p \to {\bf \Upsilon}^*}\text{grad}_\mathcal{M} h_g({\bf \Upsilon}^p) )\end{array}\right\|\\&+\left\|\begin{array}{l}\text{Tran}_{{\bf \Upsilon}^p \to {\bf \Upsilon}^*}\text{grad}_\mathcal{M} f({\bf \Upsilon}^p)\\+\sum_{g=1}^G {\bf \Phi}_g^{*} \text{Tran}_{{\bf \Upsilon}^p \to {\bf \Upsilon}^*}\text{grad}_\mathcal{M} h_g({\bf \Upsilon}^p) \end{array}\right\| . \label{calbignorm}
\end{aligned}
\end{equation}
To do with the first term in (\ref{calbignorm}), we utilize the \textit{Lemma \ref{lemmaAPC1}} in the following. 

\begin{lemma} \label{lemmaAPC1}
Given a Riemannian manifold $\mathcal{M}$, a function $f$ (continuous differentiable), and a point ${\bf \Upsilon}\in\mathcal{M}$, if ${\bf \Upsilon}^0,{\bf \Upsilon}^1,{\bf \Upsilon}^2,...$ is a sequence of points in a normal neighborhood of ${\bf \Upsilon}$ and convergent to ${\bf \Upsilon}^*$, then the following holds,
\begin{equation}
    \lim_{p \to \infty} \| \text{Tran}_{{\bf \Upsilon}^p \to {\bf \Upsilon}^*}\text{grad}_\mathcal{M} f({\bf \Upsilon}^p) - \text{grad}f({\bf \Upsilon})  \|=0.
\end{equation}
\end{lemma}

\textit{Proof}: See \textit{Lemma A.2} in \cite{liu2020simple}. $\hfill\blacksquare$

Based on the lemma \ref{lemmaAPC1}, the first term in (\ref{calbignorm}) vanishes in the $\lim p \to \infty$ since ${\bf \Upsilon}^p \to {\bf \Upsilon}^*$. For the second term in (\ref{calbignorm}), it can be written as,
\begin{subequations}
\begin{align}
    &\left\|\begin{array}{l}\text{Tran}_{{\bf \Upsilon}^p \to {\bf \Upsilon}^*}\text{grad}_\mathcal{M} f({\bf \Upsilon}^p)\\+\sum_{g=1}^G {\bf \Phi}_g^{*} \text{Tran}_{{\bf \Upsilon}^p \to {\bf \Upsilon}^*}\text{grad}_\mathcal{M} h_g({\bf \Upsilon}^p) \end{array}\right\| \\&=\|\text{grad}_\mathcal{M} f({\bf \Upsilon}^p)+\textstyle\sum_{g=1}^G {\bf \Phi}_g^{*} \text{grad}_\mathcal{M} h_g({\bf \Upsilon}^p) \|\label{usehhsgdhs1}\\&\le\|\text{grad}_\mathcal{M} f({\bf \Upsilon}^p)+\textstyle\sum_{g=1}^G {\bf \Phi}_g^{p} \text{grad}_\mathcal{M} h_g({\bf \Upsilon}^p) \|\notag\\&\quad\quad+\left\|\textstyle\sum_{g=1}^G ({\bf \Phi}_g^{*}-{\bf \Phi}_g^{p}) \text{grad}_\mathcal{M} h_g({\bf \Upsilon}^p) \right\|,\label{usehhsgdhs2}
\end{align}    
\end{subequations}    
where the equality (\ref{usehhsgdhs1}) follows from the isometry of and linearity of the vector transport operation, and the inequality (\ref{usehhsgdhs2}) follows from the triangle inequality. Here, the first term in (\ref{usehhsgdhs2}) vanishes in the limit because it tends to zero based on the proof in (\ref{finalproofth1}), and the second term in (\ref{usehhsgdhs2}) attains arbitrarily small values for large $p$ as norms
of gradients are bounded in a neighbourhood of ${\bf \Upsilon}^*$ and by definition of ${\bf \Phi}_g^{*}$. Since ${\bf T}$ is independent of $p$, we conclude that $\|\bf T\| = 0$. Therefore, ${\bf \Upsilon}^*$ satisfies KKT conditions \cite{boyd2004convex}.

On the other hand, if $\mathcal{E}^p$ is unbounded, define ${\bf \tilde \Phi}_g^p={\bf \Phi}_g^p \oslash \mathcal{E}^p$, we have,
\begin{equation}
    \left\| \frac{1}{\mathcal{E}^p} \text{grad}_{\mathcal{M}}f({\bf \Upsilon}^p)+ \sum_{g=1}^G {\bf \tilde \Phi}_g^p\text{grad}_{\mathcal{M}}h_g({\bf \Upsilon}^p)\right\|={{\bf T}}\oslash \mathcal{E}^p.
\end{equation}
As $\|{\bf \tilde \Phi}_g^p\|_{\infty}$ on the left-hand side are bounded in $[-1,1]$, and by the definition of $\mathcal{E}^p$, ${\bf \tilde \Phi}_g^p$ has a nonzero element limit point, denote it as ${\bf \tilde \Phi}_g^*$. By taking the limit in $p$, we can obtain,
\begin{equation}
     \left\| \sum_{g=1}^G {\bf \tilde \Phi}_g^*\text{grad}_{\mathcal{M}}h_g({\bf \Upsilon}^*)\right\|=0.
\end{equation}
This contradicts the LICQ condition in (\ref{inequavutLIC}). Hence, the case where $\mathcal{E}^p$ is unbounded does not occur. Therefore, we are left with the case where $\mathcal{E}^p$ is bounded, and for this case, we have already shown that the limit point ${\bf \Upsilon}^*$ satisfies KKT condition, which also implies that ${\bf \Upsilon}^*$ is a stationary point of the problem in (\ref{reformualteobnew4}). This completes the proof.  $\hfill\blacksquare$

\end{appendices} 

\bibliographystyle{IEEEtran}
\bibliography{Bibliography}

\vfill

\end{document}